\documentclass[aps,prb,reprint,superscriptaddress,amsfonts,amssymb,amsmath,floatfix,footinbib,twocolumn,notitlepage]{revtex4-2}
\usepackage{graphicx}
\usepackage[bookmarks=true,bookmarksnumbered=true,colorlinks=true,allcolors=blue,pdfdisplaydoctitle=true]{hyperref}
\hypersetup{
  pdftitle = {Characterization of two fast-turnaround dry dilution refrigerators for scanning probe microscopy},
  pdfauthor = {Mark E. Barber, Yifan Li, Jared Gibson, Jiachen Yu, Zhanzhi Jiang, Yuwen Hu, Zhurun Ji, Nabhanila Nandi, Jesse C. Hoke, Logan Bishop-Van Horn, Gilbert R. Arias, Dale J. Van Harlingen, Kathryn A. Moler, Zhi-Xun Shen, Angela Kou, and Benjamin E. Feldman},
  pdfcreator = {},
  pdfproducer = {},
}
\usepackage{array}
\usepackage{multirow}
\usepackage{siunitx,booktabs}

\usepackage{etoolbox}
\patchcmd{\section}{\centering}{\raggedright\sffamily}{}{}
\patchcmd{\subsection}{\centering}{\raggedright\sffamily}{}{}

\renewcommand{\thefigure}{{\bfseries \arabic{figure}}}
\makeatletter
\renewcommand{\@caption@fignum@sep}{{\bfseries.} }
\makeatother

\begin{document}
\pdfsuppresswarningpagegroup=1

\title{Characterization of two fast-turnaround dry dilution refrigerators for scanning probe microscopy}

\author{Mark E. Barber}
\email{mebarber@stanford.edu}
\affiliation{Stanford Institute for Materials and Energy Sciences, SLAC National Accelerator Laboratory, Menlo Park, CA 94025, USA.}
\affiliation{Department of Applied Physics, Stanford University, Stanford, CA 94305, USA.}
\affiliation{Department of Physics, Stanford University, Stanford, CA 94305, USA.}
\affiliation{Geballe Laboratory of Advanced Materials, Stanford University, Stanford, CA 94305, USA.\looseness=-1}

\author{Yifan Li}
\affiliation{Stanford Institute for Materials and Energy Sciences, SLAC National Accelerator Laboratory, Menlo Park, CA 94025, USA.}
\affiliation{Department of Physics, Stanford University, Stanford, CA 94305, USA.}
\affiliation{Geballe Laboratory of Advanced Materials, Stanford University, Stanford, CA 94305, USA.\looseness=-1}

\author{Jared Gibson}
\affiliation{Department of Physics, University of Illinois at Urbana-Champaign, Urbana, IL 61801, USA.\looseness=-1}
\affiliation{Materials Research Laboratory, University of Illinois at Urbana-Champaign, Urbana, IL 61801, USA.\looseness=-1}

\author{Jiachen Yu}
\affiliation{Stanford Institute for Materials and Energy Sciences, SLAC National Accelerator Laboratory, Menlo Park, CA 94025, USA.}
\affiliation{Department of Applied Physics, Stanford University, Stanford, CA 94305, USA.}
\affiliation{Geballe Laboratory of Advanced Materials, Stanford University, Stanford, CA 94305, USA.\looseness=-1}

\author{Zhanzhi Jiang}
\affiliation{Department of Physics, University of Illinois at Urbana-Champaign, Urbana, IL 61801, USA.\looseness=-1}
\affiliation{Materials Research Laboratory, University of Illinois at Urbana-Champaign, Urbana, IL 61801, USA.\looseness=-1}

\author{Yuwen Hu}
\affiliation{Stanford Institute for Materials and Energy Sciences, SLAC National Accelerator Laboratory, Menlo Park, CA 94025, USA.}
\affiliation{Department of Physics, Stanford University, Stanford, CA 94305, USA.}
\affiliation{Geballe Laboratory of Advanced Materials, Stanford University, Stanford, CA 94305, USA.\looseness=-1}

\author{Zhurun Ji}
\affiliation{Department of Applied Physics, Stanford University, Stanford, CA 94305, USA.}
\affiliation{Department of Physics, Stanford University, Stanford, CA 94305, USA.}
\affiliation{Geballe Laboratory of Advanced Materials, Stanford University, Stanford, CA 94305, USA.\looseness=-1}

\author{Nabhanila Nandi}
\affiliation{Stanford Institute for Materials and Energy Sciences, SLAC National Accelerator Laboratory, Menlo Park, CA 94025, USA.}
\affiliation{Department of Physics, Stanford University, Stanford, CA 94305, USA.}
\affiliation{Geballe Laboratory of Advanced Materials, Stanford University, Stanford, CA 94305, USA.\looseness=-1}

\author{Jesse C. Hoke}
\affiliation{Stanford Institute for Materials and Energy Sciences, SLAC National Accelerator Laboratory, Menlo Park, CA 94025, USA.}
\affiliation{Department of Physics, Stanford University, Stanford, CA 94305, USA.}
\affiliation{Geballe Laboratory of Advanced Materials, Stanford University, Stanford, CA 94305, USA.\looseness=-1}

\author{Logan Bishop-Van Horn}
\affiliation{Stanford Institute for Materials and Energy Sciences, SLAC National Accelerator Laboratory, Menlo Park, CA 94025, USA.}
\affiliation{Department of Physics, Stanford University, Stanford, CA 94305, USA.}
\affiliation{Geballe Laboratory of Advanced Materials, Stanford University, Stanford, CA 94305, USA.\looseness=-1}

\author{Gilbert R. Arias}
\affiliation{Department of Physics, University of Illinois at Urbana-Champaign, Urbana, IL 61801, USA.\looseness=-1}
\affiliation{Materials Research Laboratory, University of Illinois at Urbana-Champaign, Urbana, IL 61801, USA.\looseness=-1}

\author{Dale J. Van Harlingen}
\affiliation{Department of Physics, University of Illinois at Urbana-Champaign, Urbana, IL 61801, USA.\looseness=-1}
\affiliation{Materials Research Laboratory, University of Illinois at Urbana-Champaign, Urbana, IL 61801, USA.\looseness=-1}

\author{Kathryn A. Moler}
\affiliation{Stanford Institute for Materials and Energy Sciences, SLAC National Accelerator Laboratory, Menlo Park, CA 94025, USA.}
\affiliation{Department of Applied Physics, Stanford University, Stanford, CA 94305, USA.}
\affiliation{Department of Physics, Stanford University, Stanford, CA 94305, USA.}
\affiliation{Geballe Laboratory of Advanced Materials, Stanford University, Stanford, CA 94305, USA.\looseness=-1}

\author{Zhi-Xun Shen}
\affiliation{Stanford Institute for Materials and Energy Sciences, SLAC National Accelerator Laboratory, Menlo Park, CA 94025, USA.}
\affiliation{Department of Applied Physics, Stanford University, Stanford, CA 94305, USA.}
\affiliation{Department of Physics, Stanford University, Stanford, CA 94305, USA.}
\affiliation{Geballe Laboratory of Advanced Materials, Stanford University, Stanford, CA 94305, USA.\looseness=-1}

\author{Angela Kou}
\affiliation{Department of Physics, University of Illinois at Urbana-Champaign, Urbana, IL 61801, USA.\looseness=-1}
\affiliation{Materials Research Laboratory, University of Illinois at Urbana-Champaign, Urbana, IL 61801, USA.\looseness=-1}

\author{Benjamin E. Feldman}
\email{bef@stanford.edu}
\affiliation{Stanford Institute for Materials and Energy Sciences, SLAC National Accelerator Laboratory, Menlo Park, CA 94025, USA.}
\affiliation{Department of Physics, Stanford University, Stanford, CA 94305, USA.}
\affiliation{Geballe Laboratory of Advanced Materials, Stanford University, Stanford, CA 94305, USA.\looseness=-1}

\date{\today}

\begin{abstract}
Low-temperature scanning probe microscopes (SPMs) are critical for the study of quantum materials and quantum information science.
Due to the rising costs of helium, cryogen-free cryostats have become increasingly desirable.
However, they typically suffer from comparatively worse vibrations than cryogen-based systems, necessitating the understanding and mitigation of vibrations for SPM applications.
Here we demonstrate the construction of two cryogen-free dilution refrigerator SPMs with minimal modifications to the factory default and we systematically characterize their vibrational performance.
We measure the absolute vibrations at the microscope stage with geophones, and use both microwave impedance microscopy and a scanning single electron transistor to independently measure tip-sample vibrations.
Additionally, we implement customized filtering and thermal anchoring schemes, and characterize the cooling power at the scanning stage and the tip electron temperature.
This work serves as a reference to researchers interested in cryogen-free SPMs, as such characterization is not standardized in the literature or available from manufacturers.
\end{abstract}

\maketitle

\section{Introduction}\label{sec_introduction}

The invention of the $^3$He/$^4$He dilution refrigerator (DR) marked a pivotal milestone for experiments at ultra-low temperatures.
By utilizing the enthalpy of mixing in the transition from a $^3$He-concentrated phase to a $^3$He-dilute phase, DRs achieve continuous refrigeration down to the milli-Kelvin regime.
Since the pioneering work of Das et al.\ in 1965~\cite{Das1965_LT}, which achieved a temperature of 0.22~K, subsequent design improvements have facilitated base temperatures below 10~mK~\cite{Zu2022_Cryogenics}, pushing the boundaries of accessible temperatures for scientific exploration, and making DRs indispensable tools in modern quantum materials and quantum information research.

The operation of DRs relies on a stable pre-cooling environment in the Kelvin regime to liquefy the helium mixture and maintain $^3$He circulation.
For cryogen-based (``wet'') DRs, the pre-cooling environment is achieved by immersing the system in a liquid $^4$He bath at $\sim$4~K and pumping on a small volume of the $^4$He, known as a 1~K pot.
In wet systems, the liquid $^4$He bath typically boils off over days-long timescales, requiring the users to regularly refill the helium.
Additionally, due to global helium shortages and increasing costs of liquid helium, the operation of wet DRs has become increasingly unappealing.
An alternative route to achieve low temperatures is to use cryogen-free (``dry'') DRs, where a closed-cycle pulse tube refrigerator pre-cools the cryostat to $\sim$4~K and a Joule–Thomson stage condenses the circulating $^3$He.
This alleviates the need for a liquid helium supply, but at the cost of increased mechanical vibrations from the pulse tube~\cite{Tomaru2004_Cryogenics,Chijioke2010_Cryogenics,Olivieri2017_NIMPRA,D'Addabbo2018_Cryogenics}.
These vibrations make dry DRs a challenging platform for scanning probe experiments, where ultra-low vibrations are essential.

The conventional wisdom for achieving a low vibration scanning probe microscope (SPM) is to make the microscope structure as stiff as possible and isolate it from vibration sources using a damped spring stage, such that the cut-off frequency of the spring stage precludes exciting vibrational modes of the microscope~\cite{Song2010_RSI}.
For a wet DR this can mean floating the whole cryostat on air springs to isolate it from a noisy environment, but for a dry DR where the dominant source of vibrations is the pulse tube cooler, the vibration isolation has to be internal to the DR.

One possible isolation scheme is to attach the scanning stage to a large mass and suspend it on springs below the mixing chamber~\cite{Hudson1996_CJP,deWit2019_RSI}.
This method has been used for scanning gate microscopy~\cite{Pelliccione2013_RSI} and near-field scanning microwave microscopy~\cite{Geaney2019_SciRep}, using 6 kg and 5 kg masses respectively.
By additionally suspending the whole still stage and dilution unit on springs, and using three 5 kg spring stages in series below the mixing chamber, scanning tunneling microscopy and magnetic resonance force microscopy have been demonstrated~\cite{denHaan2014_RSI}.
However, these designs require heavy modification to the DR, and the spring stages severely limit the cooling power available for the microscope~\cite{Pelliccione2013_RSI}.
Furthermore, if the DR incorporates a magnet, this can limit the space available for spring extension below the mixing chamber, and the whole microscope and suspended stage must be constructed from non-magnetic components to prevent unwanted motion in applied magnetic field~\cite{Pelliccione2013_RSI}.

The combination of low temperature and high magnetic field is often advantageous for the investigation of quantum materials.
This motivates an alternative route to implement the isolation between the pulse tube cooler and the DR: rigidly mounting the microscope to the mixing chamber.
Many modern dry DRs have this type of isolation installed by the manufacturer, and scanning systems in such cryostats have been implemented for scanning superconducting quantum interference device (SQUID) microscopy~\cite{Low2021_RSI} and for scanning nitrogen-vacancy magnetometry~\cite{Scheidegger2022_APL}.
An additional benefit of this scheme is that a fast-turnaround sample loader can be used for the microscope to drastically improve sample cool-down and warm-up times from days, in a system with a magnet installed, to hours.
Fast-turnaround sample loaders have been used to implement scanning SQUID-on-tip microscopy~\cite{Zhou2023_RSI}, and scanning microwave impedance microscopy~\cite{Cao2023_RSI}.

Despite the number of demonstrated microscopes, the large variance in calibration methods and vibration isolation strategies makes it challenging to derive a definitive estimate of the expected impact a dry DR has on scanning performance based solely on this existing body of research~\cite{Pelliccione2013_RSI,denHaan2014_RSI,Low2021_RSI,Zhou2023_RSI}.
Therefore, systematically characterizing the vibrations of modern dry DR systems is crucial for future high resolution SPMs.

In this paper, we describe two fast-turnaround scanning probe microscopy systems implemented in commercial bottom-loading dry DRs with minimal customization to the factory-default systems.
We characterize the absolute magnitude of vibrations in the bottom-loading sample pucks using geophone measurements, both with and without the pulse tube running.
We show that details of the cryostat resonances and their overlap with pulse tube harmonics dictate which modes contribute most prominently to the vibrational noise.
We further construct two SPMs based on stacks of commercial positioners from Attocube rigidly mounted in the sample puck.
We determine the relative tip-sample vibrations at low temperatures using scanning microwave impedance microscopy (MIM) and a scanning single-electron transistor (SET), using the method of characterizing the noise spectral density in a known signal gradient~\cite{Schiessl2016_APL,Shperber2019_RSI,Bishop-VanHorn2019_RSI,Low2021_RSI,Zhou2023_RSI}.
Additionally, we discuss the cooling power available for the microscope and describe specially designed filtering schemes that allow us to achieve an electron temperature of $<$30~mK in the scanning probe, characterized by Coulomb blockade thermometry with a scanning SET.
We end with a discussion of possible ways to further improve the performance of these systems.

\section{Cryostat and microscope design}\label{sec_cryostat_and_microscope_design}

\begin{figure*}[t]
\centering
\includegraphics{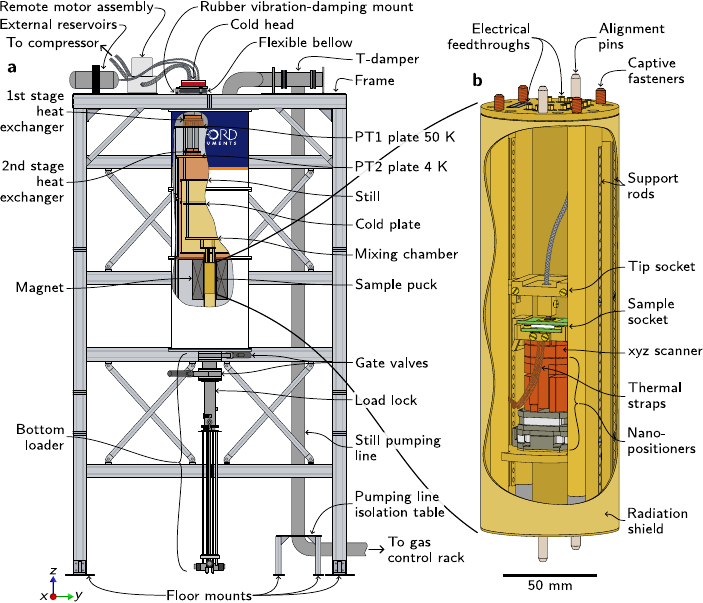}
\caption{\textbf{a}~Schematic overview of the dilution refrigerator design including the factory installed vibration isolation measures.
\textbf{b}~Schematic of a bottom loading sample puck housing the scanning probe microscope for system 1.
See Supplementary Information for photographs and additional details of both systems.}\label{fig1}
\end{figure*}

Our SPMs are implemented in Oxford Instruments ProteoxMX DRs with base temperatures of 8~mK at the mixing chamber for system 1 and $<$10~mK for system 2.
The systems incorporate superconducting vector magnets with a 6~T vertical solenoid and 1~T in-plane coils for both horizontal directions.
Pre-cooling in both systems is provided by a PT415 pulse tube cryocooler from Cryomech with remote motor option, driven by a CPA1110 compressor, which provides a cooling power of 1.35~W at 4.2~K.

The pulse tube achieves cooling though the cyclic expansion of compressed helium gas.
The remote motor turns a rotary valve that alternately connects the pulse tube inlet to the high and low pressure sides of the compressor, oscillating the pressure between 2 and 0.6~MPa at a frequency of 1.4~Hz.
The surge of gas as the valves open and close leads to the characteristic chirping sound of a pulse tube cooler and induces mechanical vibrations as parts of the pulse tube cooler expand and contract.

Several vibration isolation measures are implemented by Oxford Instruments to limit the mechanical coupling between the pulse tube and the rest of the DR; see Fig.~\ref{fig1}a for an overview.
The cold head of the pulse tube is sealed to the outer vacuum can of the DR using a flexible vacuum bellow and the weight of the cold head is supported on rubber dampers.
The cold ends of the two-stage cryocooler are joined to the respective plates, PT1 and PT2, in the cryostat through soft flexible copper braids.
The remote motor and external reservoirs for the two cooling stages are mounted directly on the top of the cryostat frame.
The top plate of the frame weighs about 100~kg, and the frame has mounting plates for anchoring to the laboratory floor.
For system 1, the remote motor of the pulse tube is driven by a linear microstepping drive from Precision Motion Controls with 50,000 steps per revolution to help reduce vibrations that can come from the jerking motion of a stepper motor and to limit the electromagnetic interference created by rapid changes in current through the magnet windings.
The compressor for the pulse tube cooler in both systems is housed in an adjacent pump room.

Additional measures are taken to limit vibrations from operation of the DR.
For system 1, the gas handling system that circulates the $^3$He is housed in the adjacent pump room with the compressor.
For the other system it is next to the cryostat, but to prevent additional vibrations due to the pumps in the gas handling system, both systems utilize an additional isolation table for the still pumping line that can be anchored to the floor.
The still pumping line also incorporates a flexible bellow type, T-damper on the top of fridge to further reduce vibrational coupling.

The cool-down time for these systems with the magnet installed is approximately 3 days due to the limited cooling power of the pulse tube cooler compared to that of liquid helium.
To facilitate faster turnaround times for exchanging the sample, a bottom loading mechanism allows the sample to be exchanged without the need to warm the entire cryostat to room temperature~\cite{Batey2014_Cryogenics}.
This has the added benefit that the sample space can be accessed without having to dismantle the nested set of vacuum cans and radiation shields, and remove the magnet.
The sample puck, Fig.~\ref{fig1}b, that houses the microscope is loaded into the DR from a load lock under the system.
It passes through spring loaded flaps in each of the radiation shields before being bolted on to the cold finger of the mixing chamber using four captive M5 fasteners driven by piston sealed drive rods from outside the fridge.
After the puck is attached to the cold finger, the bottom loader is fully retracted from the system.
This mechanism allows the microscope to be installed on the cold finger in a manner comparable to that as if the whole fridge had been fully opened to install it.
After pumping the load lock, loading the sample puck can be accomplished by one person in less than 15~minutes, and the DR cools back down to base temperature automatically in about 8~hours.
This configuration enables efficient use of the available space and cooling power of the DR for electrical wiring and other services installed and thermally anchored in the cryostat.
Each puck has two 51-way Nano-D connectors for DC wires and up to 28 SMP connectors for coaxial high frequency connections.
With all our wiring installed, including low resistance DC wiring for driving up to eight nanopositioners, the measured cooling power available at the mixing chamber is 450~$\mu$W at 100~mK, and the expected cooling power in the sample puck is $\sim$100~$\mu$W~\cite{Batey2014_Cryogenics}.

\begin{figure*}[tp]
\centering
\includegraphics{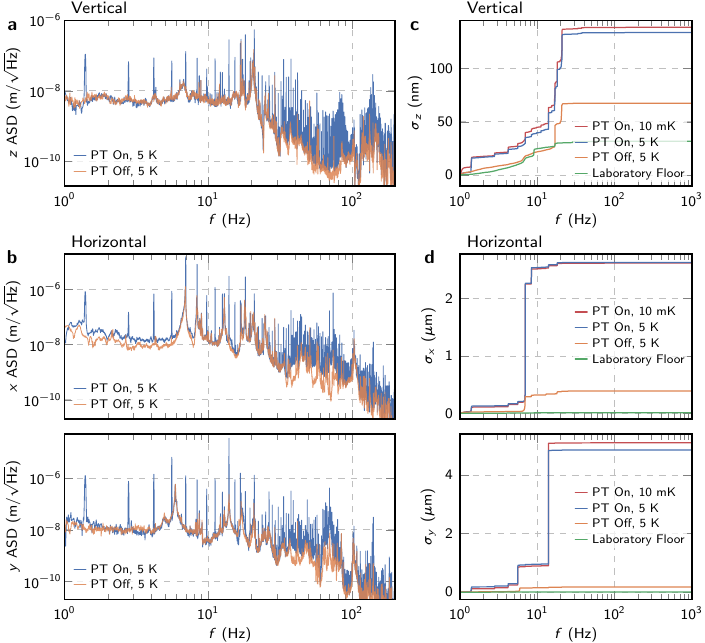}
\caption{\textbf{a}~Amplitude spectral density (ASD) of vibrations measured in the vertical direction of the sample puck mounted in the dilution fridge, with the pulse tube (PT) cooler on and off at a temperature of 5~K.
Measurements are from system 1.
\textbf{b}~Vibration measurements for the horizontal directions, $x$ and $y$, as defined in Fig.~\ref{fig1}.
\textbf{c--d}~Cumulatively integrated root mean square (RMS) vibration amplitude in the vertical direction $\sigma_{z}$ (\textbf{c}) and the horizontal directions $\sigma_{x}$ and $\sigma_{y}$ (\textbf{d}).}\label{fig2}
\end{figure*}

We have designed two SPMs which are rigidly mounted to the sample puck body as a support structure.
The microscope used for imaging and tip-sample vibrational characterization is described below (see Supplementary Information for a detailed description of the second microscope and photographs of both).
The puck has four vertical support rods with through holes for bolts, so we attached two horizontal plates across the puck, one to hold a stack of nanopositioners and the sample, and the other to hold the tip (Fig.~\ref{fig1}b).
The plates can be positioned in increments of 2.5~mm, so the microscope can accommodate different probe lengths and samples heights while ensuring the tip and sample always meet at the center of the vector magnetic field.
For nanopositioners, we use inertia-based positioners from Attocube for coarse positioning, models ANPx311/RES+/LT/HV and ANPz102/ULT/RES+/HV, and flexure-based model ANSxyz100/CuBe/LT/HV for fine scanning.
The sample socket is made from gold plated oxygen-free high thermal conductivity (OFHC) copper and clamps a removable PCB to which the sample is mounted.
A calibrated ruthenium-oxide temperature sensor is mounted to the sample socket, and the thermal connection to the cold body of the puck is provided by two clamped OFHC copper braids, with 320 strands of 70~$\mu$m wire each, as a compromise between flexibility and thermal conductivity.

\section{Vibration characterization}\label{sec_vibration_characterization}

\subsection{Absolute vibrations of the sample puck}

To understand the vibrational characteristics of each system, we first measure the absolute vibrations where the microscopes will be housed, and then directly characterize the relative tip-sample displacement in one microscope.
To measure the vibrations of the sample pucks while installed in the DR, we use geophones mounted in the puck at the position where the nanopositioners would normally be installed, see Fig.~\ref{fig1}b.
We use models GS-11D and GS-ONE LF from Geospace Technologies that both have resonance frequencies of 4.5~Hz at room temperature, and are compact enough to fit inside the sample puck.
Below this resonance frequency, the geophone sensitivity decreases quickly, setting an effective lower limit for detecting vibrations of about 1~Hz.
In three separate cool-downs, we respectively measure a vertical geophone and a horizontal geophone oriented along the $x$ and $y$ directions indicated in Fig.~\ref{fig1}, calibrating the geophone at low temperatures for each measurement~\cite{vanKann2005_RSI}.

We first detail the measurements from system 1.
The amplitude spectral density (ASD) of the sample puck displacement is plotted in Fig.~\ref{fig2}a,b, comparing the performance with the pulse tube cooler running and with it turned off while the DR is at a temperature of 5~K.
We define the total integrated root-mean-square (RMS) vibration amplitude, $\sigma_{u_{i}}$, between two frequencies $f_1$ and $f_2$ as
\begin{equation}
\sigma_{u_{i}}(f_1,f_2) = \sqrt{\int_{f_1}^{f_2} \left[ \delta u_{i} \right]^2 \mathrm{d} f},\label{eq1}
\end{equation}
where $\delta u_{i}$ is the ASD of displacement in the $u_{i}$ = $x$, $y$, or $z$ direction.
In Fig.~\ref{fig2}c,d we show the cumulatively integrated vibration from 1~Hz to 1~kHz for the cases with the pulse tube on and off at 5~K.
We also provide comparison curves with the DR running at 10~mK, and the ambient vibration amplitude measured on the laboratory floor directly under the DR with the pulse tube off.

The effect of the pulse tube cooler can be clearly seen in the vibration spectra, Fig.~\ref{fig2}a,b, appearing as a frequency comb of harmonics of the pulse tube frequency, 1.4~Hz, all the way up to the highest measured frequencies.
However, the spectral weight of most of these peaks is relatively small, as can be seen in Fig.~\ref{fig2}c,d.
In the horizontal directions the spectral density peak from the pulse tube at 1.4~Hz is around 1~$\mu$m/$\sqrt{\text{Hz}}$ and the integrated amplitude is $\sim$100~nm.
At frequencies higher than 20~Hz, the integrated amplitude of each pulse tube harmonic is generally below $\sim$20~nm.
In the vertical direction, the overall amplitudes are an order of magnitude smaller.

The dominant contributions to the total integrated amplitude appear to come from structural modes of the cryostat that are close in frequency to a harmonic of the pulse tube frequency and are therefore excited more strongly while the pulse tube is running.
These can be identified by comparing the vibration spectra with the pulse tube on and off.
The dominant contribution is at 7~Hz for the $x$ direction and 14~Hz for the $y$ direction.
In contrast, the structural mode at $\sim$5.9~Hz in the $y$ direction does not coincide with a harmonic of 1.4~Hz, and it is not more strongly excited with the pulse tube on.
However, the nearest harmonic at 5.6~Hz contributes more strongly than its neighboring harmonics.
Running the dilution unit does not significantly increase the total vibration amplitude; the pulse tube cooler still remains the dominant source of vibrations (Fig.~\ref{fig2}c,d).

In the vertical direction, the total integrated RMS vibration up to 1~kHz is 140~nm.
In the horizontal directions, it is 2.6~$\mu$m along $x$ and 5.1~$\mu$m along $y$.
All the dominant vibrations are below $\sim$20 Hz.
Comparing to a system with the same model of pulse tube cooler, but that uses a rigid bolted connection rather than flexible copper braids between the pulse tube and the DR, the influence of the pulse tube is significantly reduced, particularly for $z$ motion~\cite{Pelliccione2013_RSI}.
Specifically, Pelliccione et al.~\cite{Pelliccione2013_RSI} observed an integrated vertical vibration amplitude of 1.9~$\mu$m at the mixing chamber of their DR and 0.6~$\mu$m on their decoupled spring stage.
Horizontally, they observed $\sim$6~$\mu$m in both directions on the mixing chamber which was reduced to $\sim$1.5~$\mu$m on the spring stage, but in their work most of the remaining vibrations were at low frequencies below 10~Hz.
The flexible mount and copper braid thermalization scheme for the pulse tube in our DR is therefore helpful to minimize the vibrations in the environment of the microscope.

\begin{figure}[tp]
\centering
\includegraphics{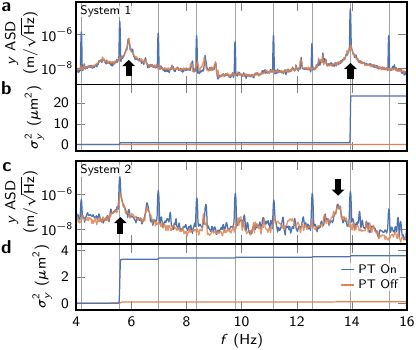}
\caption{\textbf{a}~A replication of the amplitude spectral density (ASD) of vibrations measured in the $y$ direction in the sample puck of system 1 from Fig.~\ref{fig2}, but on a finer scale from 4~Hz and 16~Hz, showing data with the pulse tube (PT) cooler on and off, at a temperature of 5~K.
Vertical lines mark harmonics of the 1.4~Hz pulse tube frequency.
Two identified structural modes of the cryostat are indicated by arrows.
\textbf{b}~Cumulatively integrated (starting from 4~Hz) vibration power spectral density (PSD).
\textbf{c--d}~Analogous vibration spectra measured in system 2 showing the vibration ASD (\textbf{c}) and integrated PSD (\textbf{d}).
Arrows indicate the shifted structural modes.}\label{fig3}
\end{figure}

To provide a second point of comparison, we conduct analogous characterization of the vibrations in the sample puck of system 2, see Supplementary Information for plots of the full vibration spectra.
In the second system, the total integrated RMS vibration amplitude from 1--200~Hz is similar in the $xy$ plane, and about a factor of two larger in $z$.
A substantially larger contribution at the 1.4~Hz pulse tube frequency is observed.
However, on further inspection of the system, it was identified that the isolation of the pulse tube from the cryostat was not correctly implemented during installation, and the cold head was rigidly connected to the top plate of the cryostat.
This may be the cause of the increased 1.4~Hz vibration (see Supplementary Information for a detailed discussion and other differences in pulse tube isolation between systems).

\begin{figure*}[tp]
\centering
\includegraphics{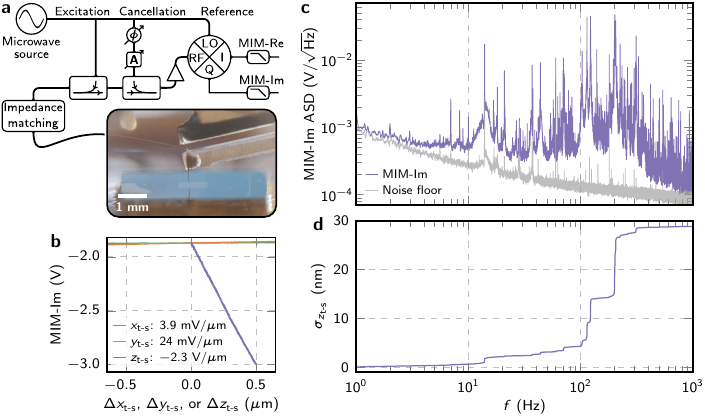}
\caption{\textbf{a}~Image of the microwave impedance microscopy (MIM) tip affixed to a tuning fork, positioned above a sample, and a schematic of the MIM readout circuit~\cite{Cui2016_RSI,Barber2022_NRP}.
\textbf{b}~Change in MIM signal MIM-Im with relative tip-sample position and linear fitting results along all three cartesian axes, at a height of 2~$\mu$m above a metallic film.
\textbf{c}~Amplitude spectral density (ASD) of the MIM noise measured at the same location as \textbf{b}, and a measurement of the MIM noise floor, see text for details.
\textbf{d}~Integrated tip-sample RMS vibration amplitude in the $z$ direction, $\sigma_{z_{\text{t-s}}}$.}\label{fig4}
\end{figure*}

While this makes it difficult to directly compare the absolute magnitude of vibrations between the two systems, we observe differences in the vibration spectra which highlight the importance of the precise frequencies of structural modes.
For example, in Fig.~\ref{fig3} we show a detailed comparison of the $y$ vibration spectrum between 4~Hz and 16~Hz for the two systems, both with the pulse tubes running and with them turned off.
The primary structural modes in this frequency range for system 1 occur at 5.9~Hz and 14~Hz (arrows in Fig.~\ref{fig3}a), as previously identified.
The presumably equivalent modes of system 2 are both at slightly lower frequencies, 5.6~Hz and 13.5~Hz (arrows in Fig.~\ref{fig3}c).
This likely reflects slight differences in the design of each cryostat.
For example, system 1 is 275~mm taller to accommodate a double gate valve on the bottom loader, which allows samples to be transferred from a glove box to the DR without exposing them to air.
In addition, the remote motor and high pressure lines are oriented in different directions in each system, and the second system is not bolted to the floor.

To compare the spectral weight between different frequencies, we plot the integrated power spectral density.
For system 1, plotted in Fig.~\ref{fig3}b, the 14~Hz mode results in the largest vibration when the pulse tube is turned on.
This is because it coincides with a harmonic of the pulse tube.
However, for the second system, the corresponding mode is not resonant with a pulse tube harmonic, and it has a smaller contribution to vibrations when the pulse tube is running (Fig.~\ref{fig3}d).
The situation is reversed for the lower frequency mode: the 5.6~Hz mode of the system 2 matches a pulse tube harmonic and has the larger contribution with the pulse tube on.
These differences emphasize the importance of considering details of the structural modes of the cryostat and optimizing design to minimize overlap between the structural modes and the pulse tube harmonics.

\subsection{Relative tip-sample vibrations}\label{subsec_relative_tip-sample_vibrations}

We now turn to direct measurements of the relative tip-sample vibrations in the SPM.
This is the important figure of merit for a scanning probe system, as it characterizes the scanning performance in the configuration used for experiments.
If the microscope is stiff enough, it can act as a `high-pass' filter and low frequency vibrations will cause the tip and sample to move together without hindering scanning performance.
If, however, there are softer modes that overlap with the absolute vibrations of the sample puck discussed above, relative tip-sample motion can be excited.

We first describe MIM measurements of the relative tip-sample vibrations of the microscope mounted in system 1 characterized above.
MIM measures the admittance between a sharp tip and the sample at microwave frequencies and is sensitive to changes of local conductivity and permittivity with nanoscale spatial resolution; see recent reviews \cite{Chu2020_ARMR} and \cite{Barber2022_NRP} for more details.
Figure~\ref{fig4}a shows the sharp metallic tip approached close to a sample, a gold film deposited on a Si/SiO$_2$ substrate, and a simplified schematic of the readout scheme for the MIM signal.
The imaginary part of the MIM signal, MIM-Im, characterizes the screening of the tip's electric field by the sample, and it diverges as the tip approaches a metallic sample.
This can be used as a sensitive measure of the tip-sample vibrations in the $z$-direction since any vibrations will appear as additional noise in the MIM signal, and the frequency dependence of the noise will contain information about the vibration spectrum.
By positioning above a spatially uniform region of the gold film, such that there is little signal change with lateral motion of the tip, the $z$ vibration spectrum can be independently obtained.
The total noise $\delta\text{MIM-Im}$ can be decomposed as
\begin{multline}
\delta\text{MIM-Im}^2 = \delta\text{MIM-Im}_{\text{intrinsic}}^2 +\\ \left( \delta z_{\text{t-s}} \frac{\partial \text{MIM-Im}}{\partial z_{\text{t-s}}} \right)^2,\label{eq2}
\end{multline}
where the intrinsic MIM noise $\delta\text{MIM-Im}_{\text{intrinsic}}$ is assumed to add independently, in quadrature, with the tip-sample vibrational noise, $\delta z_{\text{t-s}}$, and we assume the MIM-Im signal changes linearly with distance on the scale of the vibrations.

Figure~\ref{fig4}c shows the ASD of the MIM-Im noise signal with the tip positioned $\sim$2~$\mu$m above the sample, where the MIM signal is approximately linear over a few hundred nanometers in $z$ (Fig.~\ref{fig4}b), and the measured intrinsic MIM noise floor at a temperature of 10~mK.
The intrinsic MIM noise is mostly limited by the output noise of the first amplification stage, which for our measurement is a cryogenic low noise amplifier mounted to the 4~K stage of the DR.
We measure this intrinsic noise by disconnecting the excitation to the MIM tip and instead directly drive the input of the amplifier through a second directional coupler normally used for nulling the MIM signal at a particular position.
The total noise is dominated by vibrations at most frequencies.
Subtracting the power spectral densities and integrating, we obtain the integrated RMS vertical tip-sample vibration amplitude $\sigma_{z_{\text{t-s}}}$, plotted in Fig.~\ref{fig4}d.
The total integrated vibration up to 1~kHz is 29~nm.
The dominant vibrations are in the range 100--300~Hz, with the largest peaks at 114~Hz (contributing 7~nm amplitude), 122~Hz (10~nm), $\sim$200~Hz (22~nm), and 312~Hz (7~nm).
These are likely structural modes of the microscope body or nanopositioners that are being excited by the residual motion of the puck at these higher frequencies.
Many harmonics of the fundamental pulse tube frequency are visible in the vibration spectrum, but the spectral weight of most are small compared to the vibrations identified above.

The relative $xy$ vibrations between the tip and sample can be measured when the in-plane gradient of the signal is comparable to the $z$ gradient.
However, at this level of $z$ vibration, the response in $z$ remains the dominate source in the overall noise, such that we cannot reliably extract $xy$ vibrations.
To obtain a second independent measure of vertical tip-sample vibrations and address the $xy$ vibrations in our system, we use a different probe: a scanning SET~\cite{Yoo1997_Science}.

\begin{figure*}[tp]
\centering
\includegraphics{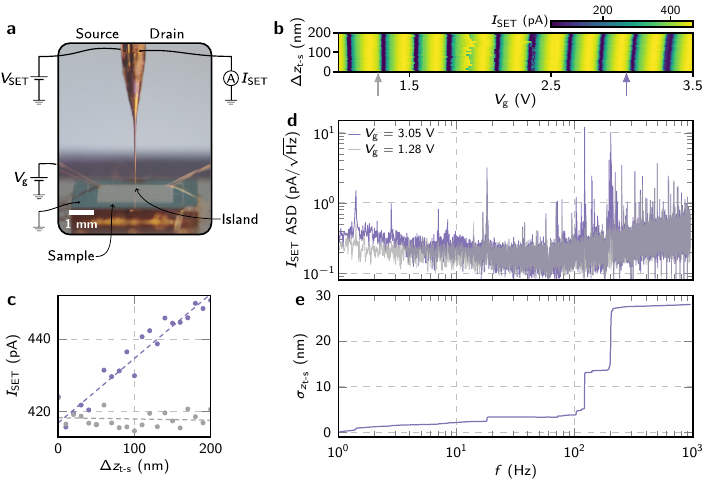}
\caption{\textbf{a}~Image of the scanning single-electron transistor (SET) above a sample, with the schematic measurement circuit overlaid and SET components labeled.
\textbf{b}~Current through the SET $I_{\text{SET}}$ as a function of gate voltage $V_{\text{g}}$ and change in tip-sample distance $\Delta z_{\text{t-s}}$ (note that $\Delta z_{\text{t-s}} = 0$ does not correspond to zero tip-sample distance).
The applied SET source-drain bias is $V_{\text{SET}}$ = 2.75~mV.
\textbf{c}~$I_{\text{SET}}$ as a function of $\Delta z_{\text{t-s}}$ at two different gate voltages.
At $V_{\text{g}}$~=~1.28~V, $I_{\text{SET}}$ is independent of $z_{\text{t-s}}$, whereas it varies linearly with $z_{\text{t-s}}$ at $V_{\text{g}}$~=~3.05~V.
\textbf{d}~Amplitude spectra density (ASD) of the filtered $I_{\text{SET}}$ at the two $V_{\text{g}}$ values in \textbf{c}, see text for details of the filtering.
\textbf{e}~Integrated tip-sample RMS vibration amplitude in the $z$ direction, $\sigma_{z_{\text{t-s}}}$, from SET measurements.}\label{fig5}
\end{figure*}

The scanning SET tip used in this work (Fig.~\ref{fig5}a) consists of source and drain electrodes separated from a metallic island by tunnel junctions, all evaporated on a pulled quartz rod as described in Ref.~\cite{Yu2022_NatPhys}.
The small size of the island, $\sim$100~nm in diameter, leads to a large charging energy $E_{\text{c}} \sim 1$~meV due to the large Coulomb repulsion from electrons already on the island.
Therefore, when the temperature is much lower than $E_{\text{c}}$, current can only flow through the SET when the energy of a state in the island matches the Fermi level of the leads~\cite{Kouwenhoven1997}.
The former depends on the electrostatic environment, giving rise to a sensitive response of SET current $I_{\text{SET}}$ to the local electrostatic potential.

To measure the relative tip-sample vibrations with the SET, we bring it close to a gold film with patterned holes deposited on a grounded Si/SiO$_2$ wafer.
This serves as a test sample and we apply a voltage $V_{\text{g}}$ to the film to electrostatically gate the SET.
The applied potential difference and the different work functions of the film, substrate and tip, lead to spatial gradients in the electrostatic potential along the $x$, $y$, and $z$ directions, which affect the measured $I_{\text{SET}}$ as the tip moves relative to the sample.
In general, assuming $I_{\text{SET}}$ varies linearly as a function of spatial position, at any given frequency $f$, the tip-sample vibrations couple into the current noise as
\begin{equation}
\delta I_{\text{SET}}^2(f) = \delta I_{\text{SET,0}}^2(f) + \left[ \nabla_\text{t-s} I_{\text{SET}} \cdot \delta\boldsymbol{r}_\text{t-s}(f) \right]^2,\label{eq:SET_noise1}
\end{equation}
where $\delta I_{\text{SET}}$ denotes the total SET current noise amplitude spectral density, $\delta I_{\text{SET,0}}$ is assumed to be uncorrelated noise that does not arise from tip-sample vibrations, such as electronic noise, $\nabla_\text{t-s} = \frac{\partial}{\partial x_\text{t-s}} \boldsymbol{e_x} + \frac{\partial}{\partial y_\text{t-s}} \boldsymbol{e_y} + \frac{\partial}{\partial z_\text{t-s}} \boldsymbol{e_z}$ is the gradient operator, and $\delta\boldsymbol{r}_\text{t-s} = \delta x \boldsymbol{e_x} + \delta y \boldsymbol{e_y} + \delta z \boldsymbol{e_z}$ is the displacement vector of tip-sample vibrations.

We first describe measurements of vibrational noise along the $z$ direction that corroborate the MIM result.
To exclude contributions from lateral vibrations, we perform our measurements above a uniform region of the gold film, where in-plane gradients are minimal, and Eq.~\ref{eq:SET_noise1} thus reduces to
\begin{equation}
\delta I_{\text{SET}}^2 = \delta I_{\text{SET,0}}^2 + \left(\delta z_\text{t-s}\frac{\partial I_\text{SET}}{\partial z_\text{t-s}}\right)^2.\label{eq:SET_noisez}
\end{equation}

In Fig.~\ref{fig5}b, we plot $I_{\text{SET}}$ as a function of $V_{\text{g}}$ as we vary the height of the tip at an approximate tip-sample distance of $\sim$1~$\mu$m.
We observe that the period of the Coulomb blockade oscillations decreases as the tip approaches the sample due to the increased tip-sample capacitance~\cite{Kouwenhoven1997}.
The corresponding change in $I_{\text{SET}}$ as a function of $\Delta z_{\text{t-s}}$ can then be utilized to measure $\delta z_{\text{t-s}}$ by comparing the current noise spectra at two distinct gate voltages with different gradients.

We first measure the SET current noise at $V_{\text{g}}$~=~3.05 V (purple arrow in Fig.~\ref{fig5}b).
At this gate voltage, $I_{\text{SET}} \sim$ 420~pA and it has a linear dependence on $\Delta z_{\text{t-s}}$ (Fig.~\ref{fig5}c).
Another current noise spectrum is taken at $V_{\text{g}}$~=~1.28~V (gray arrow in Fig.~\ref{fig5}b), which maintains the same current set point to keep non-vibrational noise contributions of the SET consistent between each measurement, but has minimal $z$ gradient.
Near this voltage, the work function difference between the tip and the sample is compensated so that the sample no longer gates the SET.
It therefore realizes a condition, termed `flat band' in the SET literature, where $I_{\text{SET}}$ is independent of $z_{\text{t-s}}$ (Fig.~\ref{fig5}c) and the SET current noise should only consist of non-vibrational contributions.

We note that in our measurement, a significant portion of the total noise spectral weight comes from non-vibrational noise, which exhibits small variations between measurements.
To prevent artifacts from the non-vibrational component in Eq.~\ref{eq:SET_noise1}, we therefore apply a series of digital notch filters to remove the non-vibrational noise at identified frequencies (see Supplementary Information for detailed discussions).
Serendipitously, the frequencies at which the non-vibrational noise dominates do not coincide with prominent vibrational noise peaks, enabling them to be reliably separated.
We show the filtered ASD spectra for current noise in Fig.~\ref{fig5}d.
Noise at the characteristic frequency of the pulse tube and its harmonics can be seen in the data, but, like the MIM measurements, its spectral weight is small.
Subtracting the power spectral densities and integrating, we obtain $\sigma_{z_{\text{t-s}}}$ as measured by the SET, plotted in Fig.~\ref{fig5}e.
The total integrated RMS relative tip-sample $z$-vibration, slightly less than 30~nm, is consistent across measurements at different $V_{\text{g}}$ (see Supplementary Information) as well as with the result from the MIM measurements.
The largest contributions also occur at similar frequencies in the 100--300~Hz range.
This supports our hypothesis that these vibrations are likely associated with modes of the microscope/nanopositioners rather than the specific scanned probe.

\begin{figure*}
\centering
\includegraphics{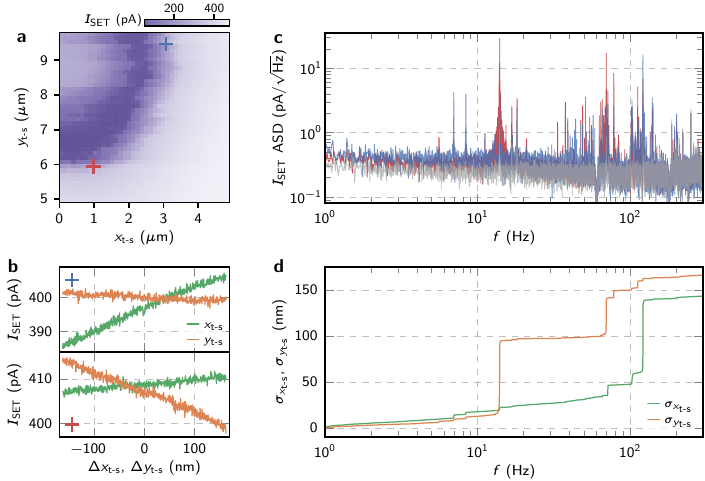}
\caption{\textbf{a}~Map of $I_\text{SET}(x,y)$ near a patterned hole in the gold film on the test sample.
The gradient is dominated by large $\partial_x I_{\text{SET}}$ ($\partial_y I_\text{SET}$) at the location marked by a blue (red) cross.
\textbf{b}~The dependence of $I_\text{SET}$ (at the flat band gate voltage) along the $x$ (green) and $y$ (orange) directions relative to the blue cross (upper panel) and the red cross (lower panel) in \textbf{a}.
\textbf{c}~Filtered amplitude spectral density ASD (see text) of $I_\text{SET}$ taken near the flat band gate voltage at $I_{\text{SET}} \sim$~400~pA.
The blue (red) curve is measured at the blue (red) cross in \textbf{a} and dominated by the vibrational noise along the $x$ ($y$) direction.
The gray trace is measured far from the hole in the film and only includes contributions from non-vibrational noise.
\textbf{d}~Integrated RMS tip-sample vibrations $\sigma_{x_{\text{t-s}}}$ and $\sigma_{y_{\text{t-s}}}$ along the $x$ and $y$ directions (green, orange) fitted from vibration measurements at multiple spatial locations (see Supplementary Information).}\label{fig6}
\end{figure*}

An analogous method can be used to determine in-plane tip-sample vibrations by measuring near the lithographically defined holes in the gold film.
The work function mismatch between the gold film and the Si/SiO$_2$ substrate, in combination with the $V_{\text{g}}$ applied to the gold film, result in an electrostatic potential gradient near the edges of the holes.
When the SET is brought close to these edges, the electrostatic potential gradient induces strong in-plane spatial modulation of $I_\text{SET}$, as illustrated by the constant-height $I_{\textrm{SET}}(x,y)$ map taken near one corner of a patterned hole in Fig.~\ref{fig6}a.
Curves with constant $I_{\text{SET}}$ in the image can be interpreted as equipotential contours at the height of the probe.

To illustrate the $xy$ vibration measurement procedure we focus on two spatial locations, marked by the blue and red crosses in Fig.~\ref{fig6}a.
The $I_{\text{SET}}$ measured along two linear trajectories through each of these points in the $x$ and $y$ directions (Fig.~\ref{fig6}b) shows a strong anisotropic response for the two locations: the blue (red) location has large $\partial I_{\text{SET}}/\partial x$ ($\partial I_{\text{SET}}/\partial y$), but small $\partial I_{\text{SET}}/\partial y$ ($\partial I_{\text{SET}}/\partial x$).
We measure $I_{\text{SET}}$ noise spectra at these two locations near the local flat band voltage, to minimize $z$-vibration contributions while ensuring the same $I_{\text{SET}} \sim$~400~pA setpoint.
The corresponding $I_{\text{SET}}$ ASD (again filtered to remove non-vibrational noise, see Supplementary Information) at these locations is shown in Fig.~\ref{fig6}c along with a measurement further away from the feature with minimal vibrational contribution.

In principle, these current noise spectra can be converted to in-plane vibrational noise using the measured gradients.
However, even though the above spectra provide a good facsimile for vibrational noise along one specific cardinal direction, perfect decoupling is challenging and in general, vibrations in both directions may contribute to the measured noise at a given location.
Therefore, we measure similar spectra at multiple spatial locations (see Supplementary Information Fig.~S7).
Each of these locations exhibits different spatial gradients along each direction. A correlation analysis shows that the major in-plane vibrational peaks only correlate to either one of the in-plane cardinal directions (see Supplementary Information), which leads to a further simplification of Eq.~\ref{eq:SET_noise1} when $x$ and $y$ vibrations are decoupled from each other:
\begin{multline}
\delta I_{\text{SET}}^2 = \delta I_{\text{SET,0}}^2 + \left(\delta x_\text{t-s}\frac{\partial I_\text{SET}}{\partial x_\text{t-s}}\right)^2 +\\ \left(\delta y_\text{t-s}\frac{\partial I_\text{SET}}{\partial y_\text{t-s}}\right)^2.\label{eq:SET_noisexy}
\end{multline}
We therefore fit the filtered $I_{\text{SET}}$ power spectral density with respect to squared gradients $(\partial I_{\text{SET}}/\partial x)^2$ and $(\partial I_{\text{SET}}/\partial y)^2$ at every frequency.
The fitted coefficients, as functions of frequency, correspond to $\delta x_{\text{t-s}}^2$ and $\delta y_{\text{t-s}}^2$ (see Eq.~\ref{eq:SET_noisexy}).
The cumulatively integrated RMS $x$- and $y$-vibration amplitudes are plotted as functions of frequency $f$ in Fig.~\ref{fig6}d.
The integrated RMS vibration amplitudes from 1~Hz to 300~Hz, $\sigma_{x_{\textrm{t-s}}}$ and $\sigma_{y_{\text{t-s}}}$, are about 140~nm and~170 nm, respectively.
These in-plane vibrations are significantly larger than the out-of-plane vibration amplitude $\sigma_{z_{\text{t-s}}}$, similar to the findings from geophone measurements.
Comparing with the absolute vibration of the sample puck measured by the geophones, there is some direct coupling to the tip-sample vibrations.
In the $x$ direction the largest vibrations identified with the geophone, $\sim$7~Hz and $\sim$8.4~Hz, have a small contribution to the relative tip-sample vibration.
In $y$, however, where the largest absolute vibration at $\sim$14~Hz was observed, there is a significant contribution to the tip-sample vibration.
The remainder of the dominant tip-sample vibrations are at higher frequencies, similar to the $z$ direction, and these are likely structural modes of the microscope body or nanopositioners.

\section{Temperature characterization}\label{sec_temperature_characterisation}

\begin{figure*}[tp]
\centering
\includegraphics{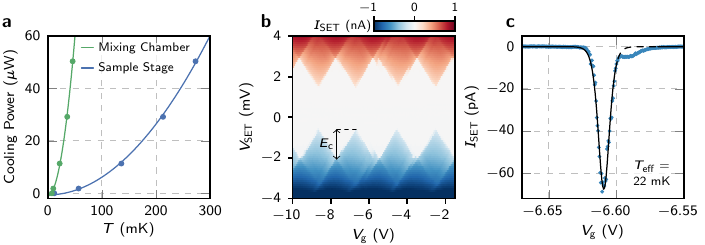}
\caption{\textbf{a}~Cooling power as a function of temperature measured at the sample stage (depicted in Fig.~\ref{fig1}b) and at the mixing chamber.
Solid curves are quadratic fits to the data.
\textbf{b}~$I_{\text{SET}}$ as a function of $V_{\text{SET}}$ and $V_{\text{g}}$ showing Coulomb diamonds at base temperature.
$E_{\text{c}}$ denotes the charging energy of the SET.
\textbf{c}~Line cut of $I_{\text{SET}}$ as a function of $V_{\text{g}}$ through the apex of one diamond at $V_{\text{SET}}$ = $-$0.6~mV (dashed line in \textbf{b}).
A best fit (solid curve) to the data using Eq.~\ref{eq4} yields an effective electron temperature of 22~mK.}\label{fig7}
\end{figure*}

The available cooling power in the sample puck is lower than that at the mixing chamber, and it is important to determine the temperature that can be obtained for the sample and probe while the microscope is in operation, as well as characterize the corresponding electron temperature.
In our microscope the sample is fixed on the nanopositioner stack, and as a consequence has a relatively weaker thermal coupling to the sample puck.
This needs to be taken into consideration to account for sources of heat at the sample stage, such as Joule heating due to large contact resistances of certain samples, and the additional heat load from the operation of the piezoelectric-based scanner.
The construction of the nanopositioners does not provide a strong thermal connection through the stack, so the sample socket is thermalized by two clamped OFHC copper braids that are flexible enough so that they do not hinder the motion of the nanopositioners.
Our microscope also utilizes two bearing-based coarse positioners for the $x$ and $y$ motion for increased rigidity, but these are made from titanium that becomes superconducting below $\sim$400~mK at zero magnetic field, additionally hindering the thermal anchoring.
We characterize this configuration by measuring the cooling power at the sample stage using a heater mounted to the sample socket and monitor the temperature with a ruthenium-oxide temperature sensor also mounted to the sample socket at zero applied magnetic field.

The resulting cooling power is shown as a function of sample socket temperature and mixing chamber temperature in Fig.~\ref{fig7}a.
The cooling power available at the sample is substantially reduced, about 6~$\mu$W at 100~mK, but as we will show below, this is still sufficient for operation of the microscope.

Heating due to the motion of the scanner is primarily due to dielectric loss in the piezoelectric actuator.
We observe that for a 30~min long raster scan covering 5~$\mu$m $\times$ 5~$\mu$m with 50~nm resolution, the temperature of the sample stage saturates at about 30~mK.
If a lower temperature is desired, a smaller scan range or a slower scan rate can be used as the power dissipated by the piezoelectric actuator scales approximately linearly with the scanning frequency and quadratically with the scan range~\cite{PI}.
The cooling power available at the sample socket could likely be increased by improving the thermal interface resistance of the clamped copper braid, for instance by e-beam welding the braid directly to the socket and the other end to a lug for attaching to the puck body.

The coarse stepping positioners are not operating while acquiring scanning data, but they do additionally dissipate power into the puck while being used to navigate around the sample.
For example, we observe that 60 steps at 100~Hz, moving the sample by $\sim$20~$\mu$m, heats the sample stage up to $\sim$120~mK and it cools back down to base temperature with a time constant of $\sim$5~min.

At these low temperatures, it can be challenging to sufficiently cool a sample or probe that must remain electrically isolated from the remainder of the cryostat, e.g. to enable electronic transport measurements.
In such cases, since the thermal resistance between the phonons and the electrons typically increases rapidly with decreasing temperature, most of the cooling has to come from the measurement leads.
This therefore necessitates that all the measurement leads are well thermalized so they provide effective cooling.
For our cryostat, all the DC wiring looms are thermally anchored between copper plates at every stage of the DR.
It is also important to prevent the measurement leads from guiding high-frequency radiation to the sample and probe which can lead to heating that raises the effective electron temperature.

\begin{figure*}[tp]
\centering
\includegraphics{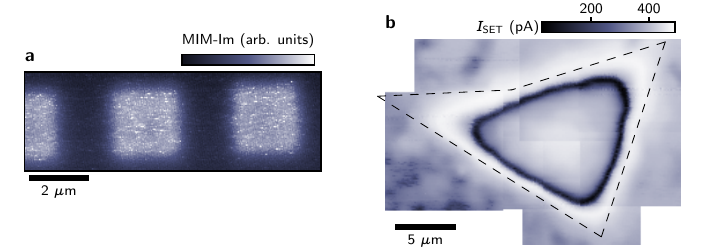}
\caption{\textbf{a}~Constant height MIM-Im scan of a 2~$\mu$m $\times$ 2~$\mu$m gold dot pattern with a 4~$\mu$m period.
\textbf{b}~Spatial map of $I_{\text{SET}}(x,y)$ in the vicinity of a patterned triangular hole feature in a gold film.
The image corresponds to multiple scans stitched together, all taken at constant height $\lesssim$1~$\mu$m above the test sample, and at fixed $V_\textrm{g}=1.2$~V.
The black dashed line marks the physical boundary of the patterned feature. }\label{fig8}
\end{figure*}

To achieve this, we integrated two different electrical filtering stages into the wiring of the DR: room-temperature radio frequency (RF) $\pi$-filters, and cryogenic RC low-pass filters.
The room temperature filter uses bolt-in style filter plates, model 52-970-206-TB0 from Spectrum Control (formerly APITech), mounted in a diecast box directly on the top of the cryostat with a short segment of shielded cable to extend the Faraday cage of the cryostat.
This filter regulates external RF noise from entering the cryostat, with a nominal 3~dB cut-off frequency of 0.65~MHz.
The cryogenic filter addresses noise originating from inside the refrigerator (e.g., triboelectric noise) and noise below the cut-off frequency of the room temperature filter.
We use a printed circuit board (PCB) based design similar to \cite{Kuemmeth2017}, but tailored to the space available on the mixing chamber of our DR, to filter out noise directly before the sample puck.
Mounting this filter on the mixing chamber minimizes any additional thermal noise from the RC filter.
The filter comprises a copper box clamping the PCB for heat sinking, and forms separate Faraday shields around three successive filtering stages.
The first stage uses a T-filter, model LFCN-225D+ from Mini-Circuits, which has a cut-off frequency of 350~MHz, then there are two successive RC filter stages with a combined cut-off frequency of 86~kHz.

To characterize the effectiveness of the filtering, we use Coulomb blockade thermometry to measure the electron temperature of a scanning SET tip~\cite{Kouwenhoven1997}.
Figure~\ref{fig7}b shows the SET current measurement as a function of $V_{\text{g}}$ and SET source-drain bias $V_{\text{SET}}$ while the DR is at base temperature.
Sharp Coulomb blockade diamonds are clearly resolved when the SET is biased above the superconducting gap.
The charging energy $E_{\text{c}}$ of the SET can be extracted from the depth of the Coulomb blockade diamonds (arrow in Fig.~\ref{fig7}b).
When the SET is biased marginally above the superconducting critical bias, the Coulomb blockade oscillations appear as sharp peaks in the SET current as a function of $V_{\text{g}}$, see Fig.~\ref{fig7}c.
The width of this peak is temperature limited when the tunneling conductance is sufficiently small such that the tunneling rate $\Gamma$ is small compared with the temperature, $\hbar\Gamma \ll k_{\text{B}} T$.
In the metallic regime, where many energy levels of the island are excited by thermal fluctuations, the line shape of an individual current peak is given by~\cite{Kouwenhoven1997}
\begin{equation}
\frac{\left| I_{\text{SET}}(V_{\text{g}}) \right| }{\left| I_{\text{SET}} \right|_{\text{max}}} \approx \cosh^{-2} \left( \frac{e \alpha \Delta V}{2.5 k_{\text{B}} T} \right),\label{eq4}
\end{equation}
where $\Delta V = V_{\text{g}} - V_0$ is the gate voltage difference to the center of the conductance peak $V_0$, and $\alpha$ characterises the ratio of the gate capacitance to the island's self capacitance which can be extracted from the aspect ratio of the Coulomb diamonds.

In Fig.~\ref{fig7}c, we fit the measured SET current peak to Eq.~\ref{eq4}, from which we obtain an effective electron temperature of $T_{\text{eff}}$ = 22~mK.
We note that the shoulder in Fig.~\ref{fig7}c is an artifact of the tip, likely related to unstable charges in or near the island.
Fitting to other current peaks also yields a similar electron temperature: $<$30~mK (see Supplementary Information).
For comparison, before the addition of the filters, the measured effective electron temperature was $T_{\text{eff}} = 120$~mK.
With the filters the extracted electron temperature is close to the reading of the calibrated temperature sensor on the sample stage, 13~mK during these measurements, indicating the filters are efficiently thermalizing the electrons on the tip.

\section{Imaging and scanner calibration}

Finally, we demonstrate the imaging capability using both MIM and SET modalities in the dry DR.
In Fig.~\ref{fig8}a we show an MIM-Im image taken at a fixed height of $\sim$100~nm above a calibration sample consisting of a 2~$\mu$m $\times$ 2~$\mu$m dot pattern with a 4~$\mu$m period, patterned in a 20~nm thick gold film on a Si/SiO$_2$ substrate.
This sample serves as a calibration of the piezoelectric scanner range, and was used to quantify the relative in-plane tip-sample vibrations determined in Sec.~\ref{subsec_relative_tip-sample_vibrations}.
We also demonstrate SET imaging of a pre-patterned triangular hole in the gold film sample described previously: maintaining a constant tip height of $\lesssim$1~$\mu$m above the sample and fixed $V_{\text{g}}$~=~1.2~V, scanning in the $xy$ plane yields the spatial map of $I_{\text{SET}}$ shown in Fig.~\ref{fig8}b.
This map is constructed by stitching together several adjacent scan frames taken with identical parameters, demonstrating the stability of the SET response, and the corresponding $I_{\text{SET}}$ modulations match well to the shape of the patterned triangular hole (dashed outline in Fig.~\ref{fig8}b).

\section{Outlook}

In summary, we describe the construction of two minimally-customized DR scanning systems, provide a systematic study of their vibrational performance, obtain an effective electron temperature of $<$30~mK in the scanning probe, and demonstrate the scanning capability of two different probes.
The factory installed vibration isolation measures to isolate the pulse tube from the DR result in sufficiently low vibration levels for certain scanning applications, setting a lower limit for the tip-sample working distance of $\sim$100~nm and a spatial resolution of several hundred nanometers.
At these vibrational levels, the microscope is suitable for studying bulk or thin film samples where the length scales of interest are on the sub-micron scale, while at ultra-low temperatures, in high magnetic field, and with multiple types of scanning probes.
Quantum anomalous Hall thin film samples could serve as an immediate candidate, with the capability of measuring current distributions and local conductivity from different conduction channels in the bulk and at the edge~\cite{Ella2019_NatNanotechnol,Allen2019_PNAS,Rodenbach2021_APLMater,Ferguson2023_NatMater}.

Our data also help identify possible schemes to further reduce the vibration levels.
The absolute vibrations of the sample puck are mostly below 20~Hz, the largest of which directly couple to the relative tip-sample vibrations.
However, the remainder of the dominant tip-sample vibrations are at higher frequencies.
The main microscope body, which currently is the standard bottom loading sample puck, could be significantly stiffened by replacing the four support rods that hold the probe and nanopositioners with a structure that has a higher bending stiffness, such as a section of a cylindrical tube~\cite{Low2021_RSI}.
If instead, the limitation of the microscope stiffness is the nanopositioners, a spring based damping could be incorporated into the sample puck, suspending a secondary structure for the microscope body.
The resonance frequency of a mass-on-spring, depends only on the extension length of the spring.
The puck is not tall enough to accommodate a very low cut-off frequency spring, but a cut-off below 10~Hz should still be possible which should provide significant damping at higher frequencies.

As well as stiffening and better isolating the microscope, additional modifications could be implemented to the DR to reduce the absolute vibration level at the sample puck.
It has been demonstrated that isolating the remote motor assembly of the pulse tube from the cryostat in a way that allows the motor assembly to swing as the high pressure lines expand and contract, rather than pushing on the DR, can further reduce vibrations~\cite{denHaan2014_RSI}.
The structural modes of the fridge identified with the geophone measurements could also be modified, for instance by adding additional bracing to the DR, to move them away from harmonics of the pulse tube frequency.
Another possible solution using a helium `battery' has been demonstrated for scanning SQUID microscopy~\cite{Franklin2023_IEEETAS}, which allows the pulse tube cooler to be turned off for a short period of time and the temporary cooling is provided by a large pumped volume of liquid helium.
However, with such a system, measurements are not necessarily as efficient as the typical cycle time is 8--10 hours, and the DR can only run without the pulse tube for a couple of hours.
An alternative to the helium `battery', using a 1~K $^4$He cooling circuit with better thermal isolation than the battery, has been shown to extend the hold time and may serve as an alternative for intermittent shutoff type measurements~\cite{Uhlig2023_Cryogenics}.

With further improvements in vibration level, we anticipate the spatial resolution of the microscope will be concurrently enhanced.
This will enable the study of mesoscopic devices and delicate nanostructures, such as van der Waals heterostructures, with spatial resolution limited by the probe, rather than vibration.

\begin{acknowledgments}
This work was supported by the QSQM, an Energy Frontier Research Center funded by the U.S. Department of Energy (DOE), Office of Science, Basic Energy Sciences (BES), under Award \# DE-SC0021238.
J.C.H. acknowledges support from the Stanford Q-FARM Quantum Science and Engineering Fellowship.
Z. Ji is supported by the Stanford Science Fellowship, and the Urbanek-Chodorow postdoctoral fellowship awards.
Z. Jiang is partially supported by the Illinois Quantum Information Science and Technology Center (IQUIST) Postdoctoral Fellowship.
Part of this work was performed at the Stanford Nano Shared Facilities (SNSF), supported by the National Science Foundation under award ECCS-2026822.
\end{acknowledgments}

\onecolumngrid
\newpage
\begin{center}
\pdfbookmark[0]{Supplementary Information for: Characterization of two fast-turnaround dry dilution refrigerators for scanning probe microscopy}{}
\textbf{\large Supplementary Information for:\texorpdfstring{\\}{~}Characterization of two fast-turnaround dry dilution refrigerators for scanning probe microscopy}\\[2.5ex]

Mark E. Barber,\textsuperscript{1,\,2,\,3,\,4,\,*} Yifan Li,\textsuperscript{1,\,3,\,4} Jared Gibson,\textsuperscript{5,\,6} Jiachen Yu,\textsuperscript{1,\,2,\,4} Zhanzhi Jiang,\textsuperscript{5,\,6} Yuwen Hu,\textsuperscript{1,\,3,\,4} Zhurun Ji,\textsuperscript{2,\,3,\,4} Nabhanila Nandi,\textsuperscript{1,\,3,\,4} Jesse C. Hoke,\textsuperscript{1,\,3,\,4} Logan Bishop-Van Horn,\textsuperscript{1,\,3,\,4} Gilbert R. Arias,\textsuperscript{5,\,6} Dale J. Van Harlingen,\textsuperscript{5,\,6} Kathryn A. Moler,\textsuperscript{1,\,2,\,3,\,4} Zhi-Xun Shen,\textsuperscript{1,\,2,\,3,\,4} Angela Kou,\textsuperscript{5,\,6} and Benjamin E. Feldman\textsuperscript{1,\,3,\,4,\,\dag}\\[1ex]

{\small \textsuperscript{1}\textit{Stanford Institute for Materials and Energy Sciences,\\
SLAC National Accelerator Laboratory, Menlo Park, CA 94025, USA.}\\
\textsuperscript{2}\textit{Department of Applied Physics, Stanford University, Stanford, CA 94305, USA.}\\
\textsuperscript{3}\textit{Department of Physics, Stanford University, Stanford, CA 94305, USA.}\\
\textsuperscript{4}\textit{Geballe Laboratory of Advanced Materials, Stanford University, Stanford, CA 94305, USA.}\\
\textsuperscript{5}\textit{Department of Physics, University of Illinois at Urbana-Champaign, Urbana, IL 61801, USA.}\\
\textsuperscript{6}\textit{Materials Research Laboratory, University of Illinois at Urbana-Champaign, Urbana, IL 61801, USA.}
}
\end{center}

\setcounter{equation}{0}
\setcounter{figure}{0}
\setcounter{table}{0}
\setcounter{page}{1}
\setcounter{section}{0}
\renewcommand{\theequation}{S\arabic{equation}}
\renewcommand{\thefigure}{{\bfseries S\arabic{figure}}}
\renewcommand{\thetable}{{\bfseries S\arabic{table}}}

\section{Detailed comparison between the two microscopes and cryostats}
\thispagestyle{empty}

\begin{figure}[bp]
\centering
\includegraphics{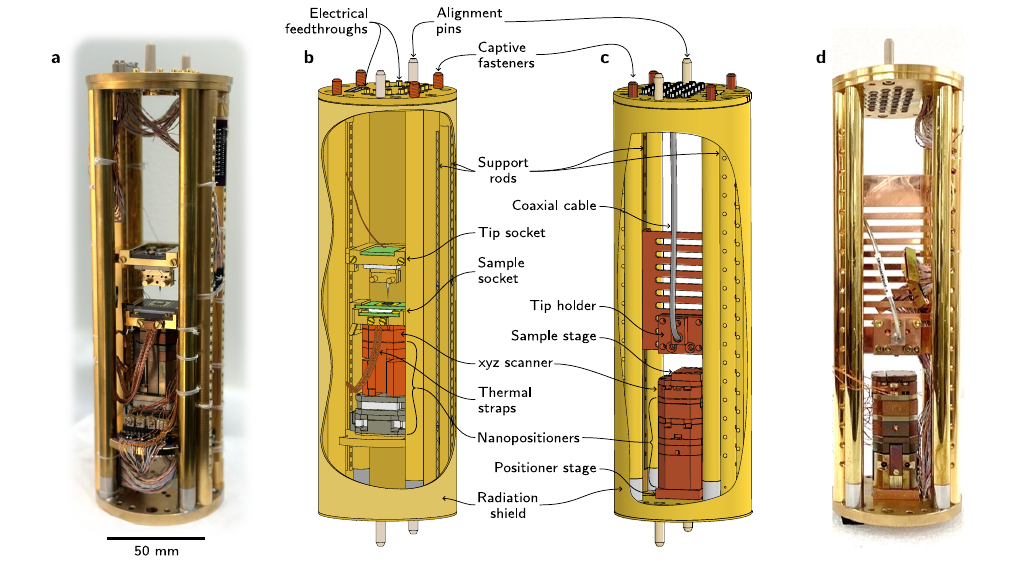}
\caption{\textbf{a--b}~Photograph and schematic of the bottom loading sample puck that houses the scanning probe microscope described in the main text, system 1.
This microscope design is suitable for microwave impedance microscopy, as shown in Fig.~1b in the main text, and scanning single-electron transistor measurements, as shown here.
\textbf{c--d}~Photograph and schematic of a second sample puck housing a microscope for scanning microwave resonator measurements, system 2.}\label{figS1}
\end{figure}

A photograph and schematic of the scanning probe microscope described in the main text, system 1, are shown in Fig.~\ref{figS1}a,b.
Here, the microscope is shown in its configuration for scanning single-electron transistor (SET) measurements, whereas in Fig.~1 in the main text, it is shown in the configuration for microwave impedance microscopy.
We have additionally designed a second microscope for scanning microwave resonator microscopy in a separate puck of a second dilution refrigerator, system 2 (Fig.~\ref{figS1}c,d).
This microscope has two 51-pin Nano-D connectors and 28 SMP connectors for DC and high-frequency signal delivery between microscope components and cryostat wiring.
Using the puck support rods, we mount a vertical bracket for the tip holder, which houses a detachable high-frequency PCB that we use to send and receive microwave signals to and from the probe.
Both the bracket and the tip holder are machined from oxygen-free high conductivity (OFHC) copper and have been polished to improve the thermal conductivity between the sample puck and tip holder.

The scanning component of this second microscope consists of three coarse positioners (one ANPz102/ULT\slash RES+/HV, and two ANPx101/ULT/RES+/HV) and one scanner (ANSxyz100/CuBe/LT/HV) for fine-positioning.
These four positioners are assembled in a stack and mounted to the bottom of the puck using a positioner stage machined from OFHC copper and polished.
A sample pedestal of polished OFHC copper is mounted on the fine-positioning scanner, to which samples can be mounted using conductive adhesive.

\begin{figure}[bp]
\centering
\includegraphics{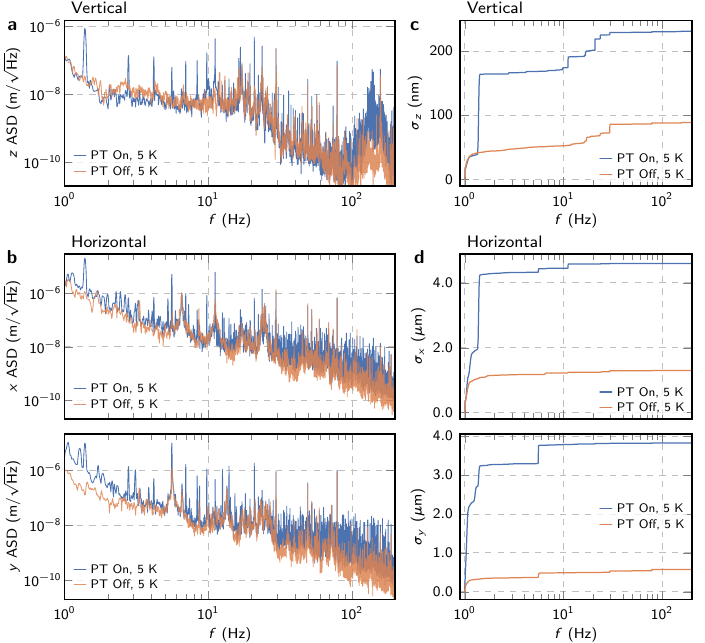}
\caption{\textbf{a}~Amplitude spectral density (ASD) of vibrations measured in the vertical direction of the sample puck mounted in system 2, with the pulse tube (PT) cooler on and off, at a temperature of 5~K.
\textbf{b}~Vibration measurements for the horizontal directions, $x$ and $y$, as defined in Fig.~1.
\textbf{c--d}~Cumulatively integrated root mean square (RMS) vibration amplitude, as defined in the main text, in the vertical direction $\sigma_{z}$ (\textbf{c}) and the horizontal directions $\sigma_{x}$ and $\sigma_{y}$ (\textbf{d}).}\label{figS2}
\end{figure}

Both systems are Oxford Instruments ProteoxMX dilution refrigerators (DR), but are configured slightly differently.
As noted in the main text the frame of system 1 is 275~mm taller to accommodate a double gate valve on the bottom loader.
The remote motor orientation, and correspondingly the direction of the high pressure lines, are different in each installation.
For system 1, the high pressure lines are parallel to the $x$ direction, as indicated in Fig.~1, before being secured to a cable tray in the ceiling.
In system 2, they are closer to the $y$ direction.
System 2 was also not bolted to the laboratory floor for the vibration measurements, whereas system 1 was.

System 1 also incorporates optical line-of-sight access vertically down the center of the cryostat into the sample puck.
The puck for this system correspondingly has only 15 SMP connectors to leave space for the opening, but for all the measurements reported here, the windows were blanked at every stage of the cryostat.

Measurements of the absolute sample puck vibrations were performed in the puck of the second microscope in a manner similar to that in system 1, as described in the main text.
The full vibration spectra are shown in Fig.~\ref{figS2}.
It was identified that the isolation of the pulse tube from the cryostat was not correctly implemented during installation for this system, and the cold head was rigidly connected to the top plate of the cryostat.
This may be the cause of the increased 1.4~Hz vibration in these spectra compared to those of system 1 presented in Fig.~2.
Qualitatively, the vibrations at higher frequencies are more similar, but the precise frequencies of structural modes are slightly different.
This may be due to the differences highlighted above, or could be due to other intrinsic variations between each DR manufactured.

\section{Vibration extraction with the SET and additional SET measurements}

Here we detail the approach and data processing used to convert the current noise measured by the SET into mechanical vibrations in the $x$, $y$, and $z$ directions.
We record 64~s long time series of SET current $I_{\text{SET}}$ at a sampling rate of 2~kHz, and Fourier transform the time series into power spectra.
We repeat each measurement multiple times and average the power spectra to obtain the presented current noise.

\subsection{Distinguishing vibrational and non-vibrational contributions}\label{subsec_corr}

In general, the current noise spectra measured by the SET include a mix of non-vibrational and vibrational contributions.
To reliably extract the mechanical vibrations, these contributions must be properly distinguished.

\begin{figure}[bp]
\centering
\includegraphics{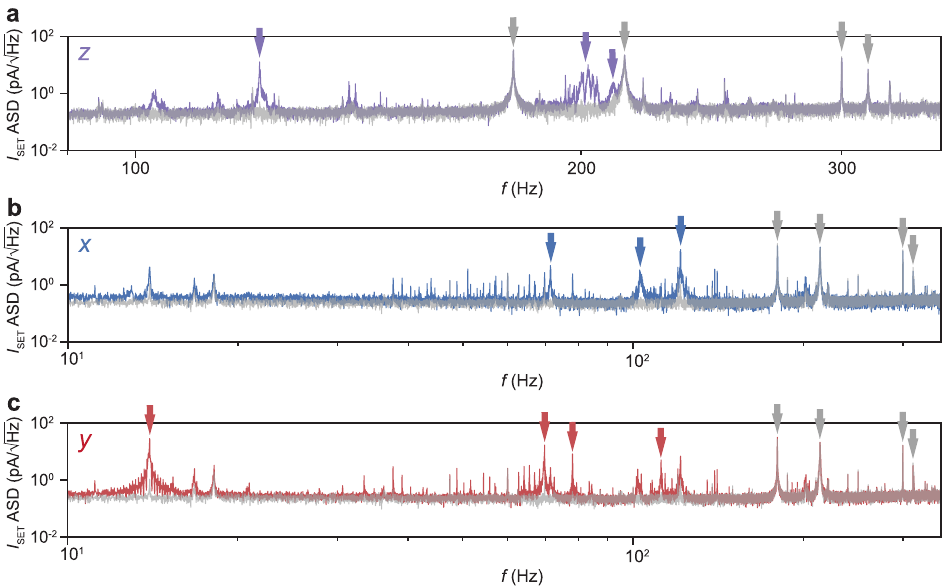}
\caption{\textbf{a}~Qualitative comparison of the amplitude spectral density (ASD) of SET current $I_\text{SET}$ comparing a zero-gradient spectrum taken at a location where the $x$, $y$, and $z$ gradients all vanish (gray curve) and a spectrum with non-zero $z$ gradient (purple curve).
The major non-vibrational noise peaks are marked with gray arrows, while the major vibrational peaks are marked with colored arrows.
\textbf{b--c}~Equivalent comparisons with spectra dominated by $x$ (\textbf{b}, blue curve) and $y$ (\textbf{c}, red curve) vibrational noise in a larger frequency range.
Note that in \textbf{b} and \textbf{c}, there is some mixing between the two in-plane directions as there was a finite gradient in both directions.
The difference in the zero-gradient spectrum in \textbf{b} and \textbf{c} relative to that in panel \textbf{a}, is due to repositioning the current preamplifier and using shorter cables.
Both spectra within a given panel were taken in the same measurement configuration and therefore can be directly compared.}\label{figS3}
\end{figure}

Qualitatively, by comparing SET current noise spectra at spatial locations with different gradients $\partial I_{\text{SET}}/\partial u_i$, with $u_i = x$, $y$, or $z$, we can attribute peaks in the spectra to either non-vibrational or vibrational contributions.
Specifically, the spectrum measured where the gradients all vanish is taken as the non-vibrational background.
We measure this above a spatially uniform region of the test sample and at a gate voltage $V_{\text{g}}$ near the `flat band' condition where the $z$ potential gradient is minimal.
The non-vibrational noise exhibits a white-like background with additional sharp peaks at several frequencies  (gray arrows, Fig.~\ref{figS3}).
In other spectra measured in the presence of non-zero spatial gradients, the vibrational contributions emerge as additional peaks in the noise on top of the non-vibrational response (colored arrows, Fig.~\ref{figS3}).
The magnitude of the vibrational peaks should scale with the gradient along their corresponding directions, while the frequency and magnitude of the non-vibrational parts of the spectra should remain unchanged in the presence of spatial gradients.
Based on this approach, we qualitatively identify the frequencies of the primary non-vibrational and vibrational peaks that contribute at least 1\% of total non-vibrational and vibrational current noise power, respectively, as listed in Table~\ref{tab:peak_freq}.

\begin{table}[t]
\centering
\renewcommand{\arraystretch}{1.3}
\setlength\tabcolsep{10pt}
\begin{tabular}{c|c!{\vrule width 1.5pt}c|c|c}
\multicolumn{2}{c!{\vrule width 1.5pt}}{\textbf{Non-vibrational}} & \multicolumn{3}{c}{\textbf{Vibrational}}\\
60 Hz harmonics & Other & $x$ & $y$ & $z$ \\
\noalign{\hrule height 1.5pt}
180 Hz & 214 Hz    & 71.1 Hz  & 14.0 Hz  & 121.3 Hz\\
300 Hz & 312.6 Hz  & 102.8 Hz & 69.7 Hz  & 198--206 Hz\\
420 Hz &           & 121.3 Hz & 78.1 Hz  & 210.2 Hz\\
       &           &          & 111.8 Hz & \\
\end{tabular}
\caption{Qualitatively identified frequencies of major non-vibrational and vibrational noise contributions.}\label{tab:peak_freq}
\end{table}

We have, however, observed drifts in peak power and frequency, which may lead to unphysical effects, such as negative power spectral density in the extracted vibration spectra.
The qualitative approach to assigning peaks may therefore become unreliable.
To provide a more quantitative and unbiased approach, we apply a correlation analysis between the current noise power of each  specific noise peak and the measured spatial gradients $\partial I_{\text{SET}}/\partial u_i$ among different datasets.

For each SET current noise spectrum, we integrate the current noise power of the major peaks centered at frequencies $f_p$ according to
\begin{equation}
P(f_p)=\int_{f_p - \Delta f}^{f_p + \Delta f}\delta I_{\text{SET}}^2(f)\text{d}f,
\end{equation}
where $\delta I_{\text{SET}}^2(f)$ is the power spectral density of $I_{\text{SET}}$, and $(f_p - \Delta f, f_p + \Delta f)$ covers the full width of each peak.
We then calculate the Pearson correlation coefficient $R_{u_i}$  ($u_i$ = $x$, $y$, $z$) of $P(f_p)$ with respect to the squared SET current gradient $(\partial I_{\mathrm{SET}}/\partial u_i)^2$:
\begin{equation}
R_{u_i}=\frac{\sum_{j}[P_j(f_p)-\overline{P_j(f_p)}]\cdot[(\partial_{u_i} I_j)^2-\overline{(\partial_{u_i} I_j)^2}]}{\sqrt{\sum_{j}[P_j(f_p)-\overline{P_j(f_p)}]^2\sum_{j}[(\partial_{u_i} I_j)^2-\overline{(\partial_{u_i} I_j)^2}]^2}},
\end{equation}
in which $j$ is the index of the current noise spectra, and $\partial_{u_i} I_j$ is shorthand notation for spatial gradient of the $j$-th spectrum.
In practice, we calculate $R_{u_i}$ and the corresponding $p$-value.
Since spurious correlations could exist due to random fluctuations, especially in the case of a small number of datasets, we apply a second metric based on the $p$-value as a measure of the validity of $R_{u_i}$, and specifically, we accept $p<0.05$ as the criterion of a valid correlation.

We summarize the results in Table~\ref{tab:xy_corr} for the $xy$ datasets.
Non-vibrational peaks, as well as $x$ and $y$ vibrational peaks can be quantitatively identified with good specificity.
In brief, the noise power of non-vibrational peaks is either not correlated with squared SET current gradients ($|R|<$~0.3), or the correlation is considered invalid ($p > 0.05$).
In contrast, the vibrational peaks are characterized by a large correlation coefficient ($|R|>$~0.8 and often close to 1) and a small $p$-value ($p < 0.05$) in the corresponding direction.

\begin{table}[t]
\centering
\renewcommand{\arraystretch}{1.3}
\setlength\tabcolsep{12pt}
\begin{tabular}{S[table-format=3.1]|S[table-format=-1.2]|S[table-format=1.4]|S[table-format=-1.2]|S[table-format=1.4]|c}
{$f_p$ (Hz)} & {$R_x$} & {$p_x$} & {$R_y$} & {$p_y$} & Type\\
\noalign{\hrule height 1.5pt}
180   & -0.13 & 0.8   &  0.21 & 0.7   & Non-vibrational \\
214   &  0.14 & 0.8   &  0.10 & 0.8   & Non-vibrational \\
300   &  0.14 & 0.8   & -0.38 & 0.4   & Non-vibrational \\
312.6 &  0.20 & 0.7   & -0.66 & 0.1   & Non-vibrational \\
420   &  0.34 & 0.5   & -0.15 & 0.7   & Non-vibrational \\
71.1  &  0.87 & 0.01  & -0.27 & 0.6   & $x$ \\
102.8 &  0.93 & 0.003 & -0.10 & 0.8   &  $x$\\
121.3 &  0.92 & 0.004 & -0.45 & 0.3   & $x$\\
14.0  & -0.41 & 0.4   &  0.95 & 0.001 & $y$\\
69.7  & -0.49 & 0.3   &  0.95 & 0.001 & $y$\\
78.1  & -0.47 & 0.3   &  0.93 & 0.002 & $y$\\
111.8 & -0.47 & 0.3   &  0.94 & 0.002 & $y$\\
\end{tabular}
\caption{Pearson correlation coefficients $R_x$, $R_y$ and the corresponding $p$-values of peaks in the SET current noise power $P(f_p)$ centered at frequencies $f_p$, with respect to $(\partial_x I)^2$ and $(\partial_y I)^2$.}\label{tab:xy_corr}
\end{table}

From Table~\ref{tab:xy_corr}, it is evident that the major $xy$ vibrational contributions are decoupled from one another, i.e. each vibrational peak is correlated to only one cardinal direction but not the other.
This indicates that the in-plane vibrations are aligned with either the $x$ or $y$ direction, and have different natural frequencies.
The cryostat frame provides a natural mechanism for symmetry breaking since it has an $x$-$y$ aspect ratio of 2.
The four support rods in the sample puck are also not spaced equal distances in $x$ and $y$, and the order of the nanopositioners in the stack affects the symmetry, so it is not unexpected that the natural frequencies are different in the $x$ and $y$ directions.
We note, however, that the $x$ and $z$ vibrations at 121.3 Hz are likely correlated, which may come from a pendulum-like mode in $x$-$z$ plane.

For the $z$ direction, we only have four datasets with different $z$ gradients, which is too few to yield reliable correlation analysis.
Therefore we assume the frequencies of non-vibrational peaks are identical for $z$ and $xy$ datasets.
This assumption does not alter our quantitative results for $z$ vibrations as we will show below.

\subsection{Digital notch filtering non-vibrational peaks}

\begin{figure}[b]
\centering
\includegraphics{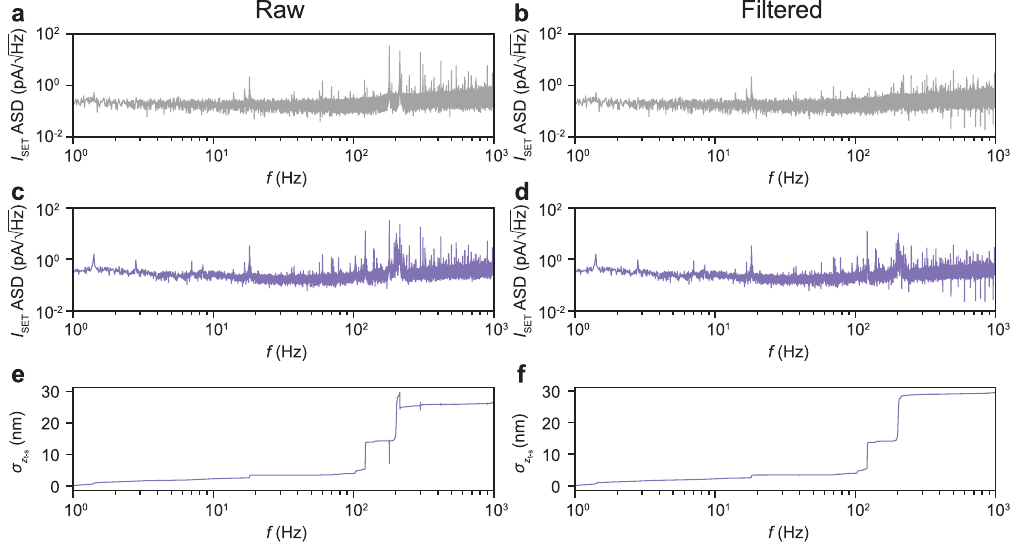}
\caption{\textbf{a}~Amplitude spectral density (ASD) of the raw SET current $I_{\text{SET}}$ measured above a spatially uniform region of the sample near the flat band voltage $V_{\text{g}}=$~1.28~V.
\textbf{b}~The spectrum from \textbf{a} after applying a series of digital notch-filters as detailed in the text.
The larger non-vibrational peaks are filtered, leaving a smoother noise background.
\textbf{c}~Raw $I_{\text{SET}}$ ASD measured near $V_\textrm{g}=$~3.05~V that includes vibrational noise in the $z$ direction.
\textbf{d}~The notch-filtered ASD spectrum of \textbf{c}.
\textbf{e}~Integrated RMS vibration amplitude $\sigma_{z_{\text{t-s}}}$ obtained from Eq.~\ref{eq:SET_znoise} based on the raw spectra shown in \textbf{a} and \textbf{c}.
The unphysical dips and jumps in the integrated $\sigma_{z_{\text{t-s}}}$ occur near the frequencies of non-vibrational peaks.
\textbf{f}~$\sigma_{z_{\text{t-s}}}$ obtained using the filtered spectra in \textbf{b} and \textbf{d}.
Panels \textbf{b}, \textbf{d}, and \textbf{f} are the same data presented in Fig.~5 in the main text.}\label{figS4}
\end{figure}

As mentioned in the main text, non-vibrational noise is dominant in the SET current noise spectra (Fig.~\ref{figS4}a,c), contributing around 90\% of the total current noise power.
This means that any variations in the non-vibrational noise over time, in frequency or power, even if small compared to the total noise, can significantly influence the apparent quantitative vibrational contribution.
Direct subtraction of the non-vibrational background according to Eq.~3 in the main text introduces unphysical effects, such as negative power spectral density in the subtracted spectrum, and is therefore inappropriate (see Fig.~\ref{figS4}e).
We detail below our procedure to account for these longer time scale drifts and remove the artifacts that would otherwise result.

The non-vibrational noise can be separated into a series of non-vibrational peaks and a smooth background.
The former contributes over 70\% of the non-vibrational noise power, and its fluctuations in frequency and noise power are large compared to the vibrational contributions.
Having previously identified that the peaks in the non-vibrational noise do not overlap in frequency with the vibrational noise, we apply a series of digital notch filters to the recorded $I_{\mathrm{SET}}$ time series to remove them prior to determining the vibrational contributions from the current noise spectra.
Each digital notch filter is a second-order infinite impulse response filter centered at non-vibrational noise frequencies. We specify the center frequency $f_0$ and bandwidth $\delta f$ of each filter in Table~\ref{tab:notch-filter}.
Note that we have also filtered smaller non-vibrational noise peaks (harmonics of 60~Hz line frequency) besides the major peaks identified in Table~\ref{tab:peak_freq} and Table~\ref{tab:xy_corr}.
After filtering, the prominent non-vibrational peaks are suppressed from the current noise spectra, leaving a smoother non-vibrational background (Fig.~\ref{figS4}b,d), and the filtering does not dramatically alter the integrated vibration amplitude we extract (Fig.~\ref{figS4}e,f).

\begin{table}[t]
\centering
\renewcommand{\arraystretch}{1.3}
\setlength\tabcolsep{5.5pt}
\begin{tabular}{c!{\vrule width 1.5pt}c c c c c c c c c c c c c c c c c c}
$f_0$ (Hz) & 60 & 120 & 180 & 240 & 300 & 360 & 420 & 480 & 540 & 600 & 660 & 720 & 780 & 840 & 900 & 960 & 214 & 312.58\\
\hline
$\delta f (\text{Hz})$ & 1 & 0.5 & 5 & 1 & 2 & 2 & 1 & 1 & 1 & 1 & 1 & 1 & 1 & 1 & 1 & 1 & 5 & 1 \\
\end{tabular}
\caption{Center frequencies $f_0$ and bandwidths $\delta f$ of the digital notch filters used in the data analysis.}
\label{tab:notch-filter}
\end{table}

\subsection{Integrating the {\fontfamily{xcmss}\fontseries{bx}\itshape\selectfont z} vibrational noise}

\begin{figure}[p]
\centering
\includegraphics{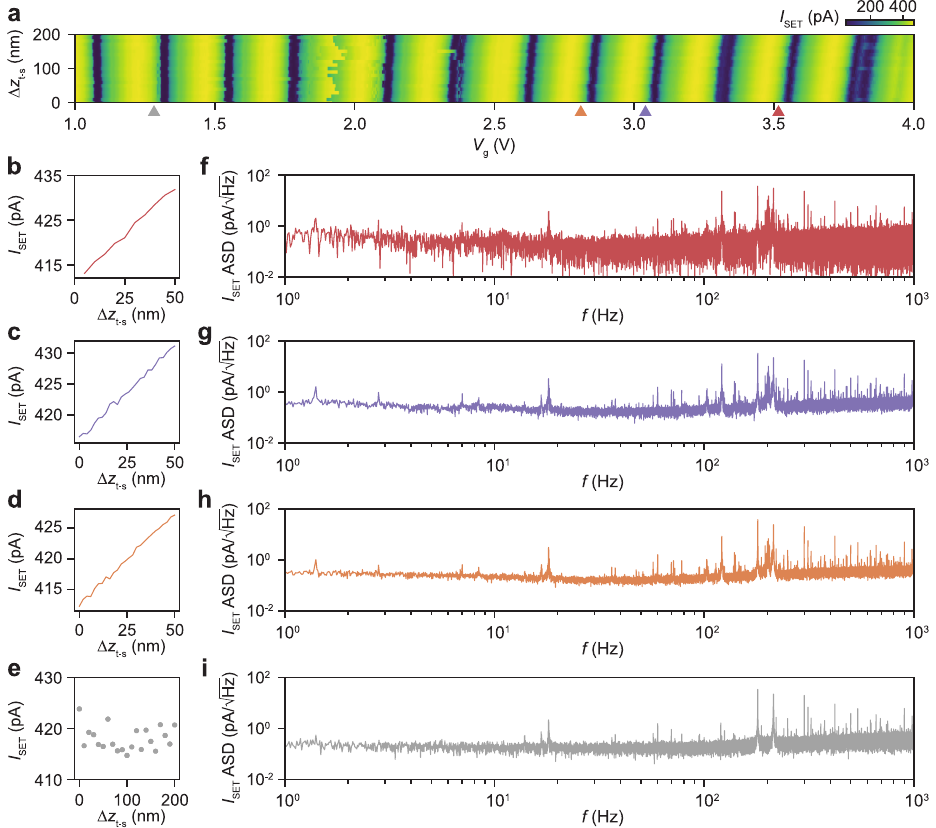}
\caption{\textbf{a}~$I_{\text{SET}}$ measured as a function of gate voltage $V_{\text{g}}$ and change in tip-sample distance $\Delta z_{\text{t-s}}$ (an extended plot of the data in Fig.~5b).
Triangles mark the $V_{\text{g}}$ at which noise spectra are measured: 1.28~V (flat band, gray), 2.82~V (orange), 3.05~V (purple) and 3.53~V (red).
\textbf{b--e}~$I_{\text{SET}}$ measured as a function of $\Delta z_{\text{t-s}}$ at $V_{\text{g}}$ marked by the triangles of the same color in \textbf{a}.
\textbf{b--d} are independently measured by recording $I_{\text{SET}}$ while the tip-sample distance is varied directly after the noise measurement, while \textbf{e} is a linecut from \textbf{a}.
\textbf{f--i}~SET current noise spectra measured at $\Delta z_{\text{t-s}} = 20$~nm and $V_{\mathrm{g}}$ indicated by the triangles of the same color.
Note that \textbf{i} is measured at the flat band voltage, and is taken as the non-vibrational reference.}\label{figS5}
\end{figure}

\begin{figure}[p]
\centering
\includegraphics{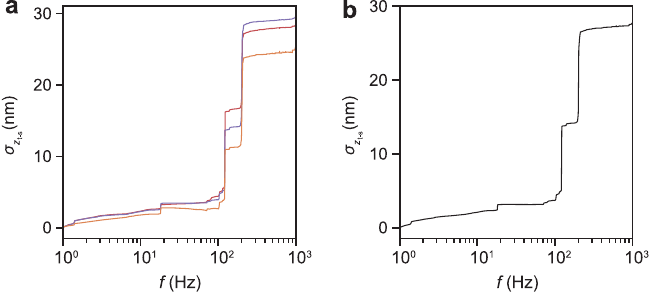}
\caption{\textbf{a}~Integrated RMS tip-sample vibration $\sigma_{z_{\text{t-s}}}$ from three independent measurements at different gate voltages $V_{\text{g}}$~=~3.53~V (red), $V_{\text{g}}$~=~3.05~V (purple), and $V_{\text{g}}$~=~2.82~V (orange).
\textbf{b}~RMS average of the three curves in \textbf{a}, resulting in a averaged total $z$ tip-sample vibration of 28~nm.}
\label{figS6}
\end{figure}

The full dataset of $z$ vibration measurements by SET consist of a flat band spectrum and three additional spectra measured at different gate voltages away from flat band: $V_{\text{g}}=$ 3.53~V, 3.05~V, and 2.82~V (Fig.~\ref{figS5}).
All four spectra are measured at the same current setpoint $I_{\text{SET}}\approx 420$~pA to minimize changes in the intrinsic electronic noise.
They are also at the same tip-sample distance and over the same uniform region of the test sample, so that $x$ and $y$ vibrational contributions are minimal.
Non-vibrational noise is filtered from the spectra as discussed above, and the $z$ tip-sample vibrations $\delta z_{\text{t-s}}$ can be extracted as a function of frequency from the differences between each non-flat band spectrum $\delta I_{\text{SET}}$ and the flat band spectrum $\delta I_{\text{SET,0}}$ according to:
\begin{equation}
\delta z_{\text{t-s}}^2(f) = \frac{\delta I_{\text{SET}}^2(f) - \delta I_{\text{SET,0}}^2(f)}{(\partial I_{\text{SET}}/\partial z_{\text{t-s}})^2}.\label{eq:SET_znoise}
\end{equation}
For each gate voltage, we determine the slope $\partial I_{\text{SET}}/\partial z_{\text{t-s}}$ by recording $I_{\text{SET}}$ as the tip height is varied over at least 50~nm (Fig.~\ref{figS5}b--e).
The total integrated root mean square (RMS) $z$-vibrations from 1~Hz to a given frequency $f$ are given by
\begin{equation}
\sigma_{z_{\text{t-s}}} = \sqrt{\int_{1\text{~Hz}}^{f} \delta z_{\text{t-s}}^2(f) df},
\end{equation}
from which we obtain three independent measurements of total $z$ vibrations up to 1~kHz of 28~$\pm~2$~nm, 29~$\pm~2$~nm, and 25~$\pm~2$~nm (Fig.~\ref{figS6}a). The primary source of uncertainty comes from the fitted slope of the gradient as a function of tip-sample displacement.
We report a RMS average value of 27~$\pm~2$~nm (Fig.~\ref{figS6}b).

\clearpage
\subsection{Fitting the {\fontfamily{xcmss}\fontseries{bx}\itshape\selectfont x} and {\fontfamily{xcmss}\fontseries{bx}\itshape\selectfont y} vibrational noise}

\begin{figure}[p]
\centering
\includegraphics{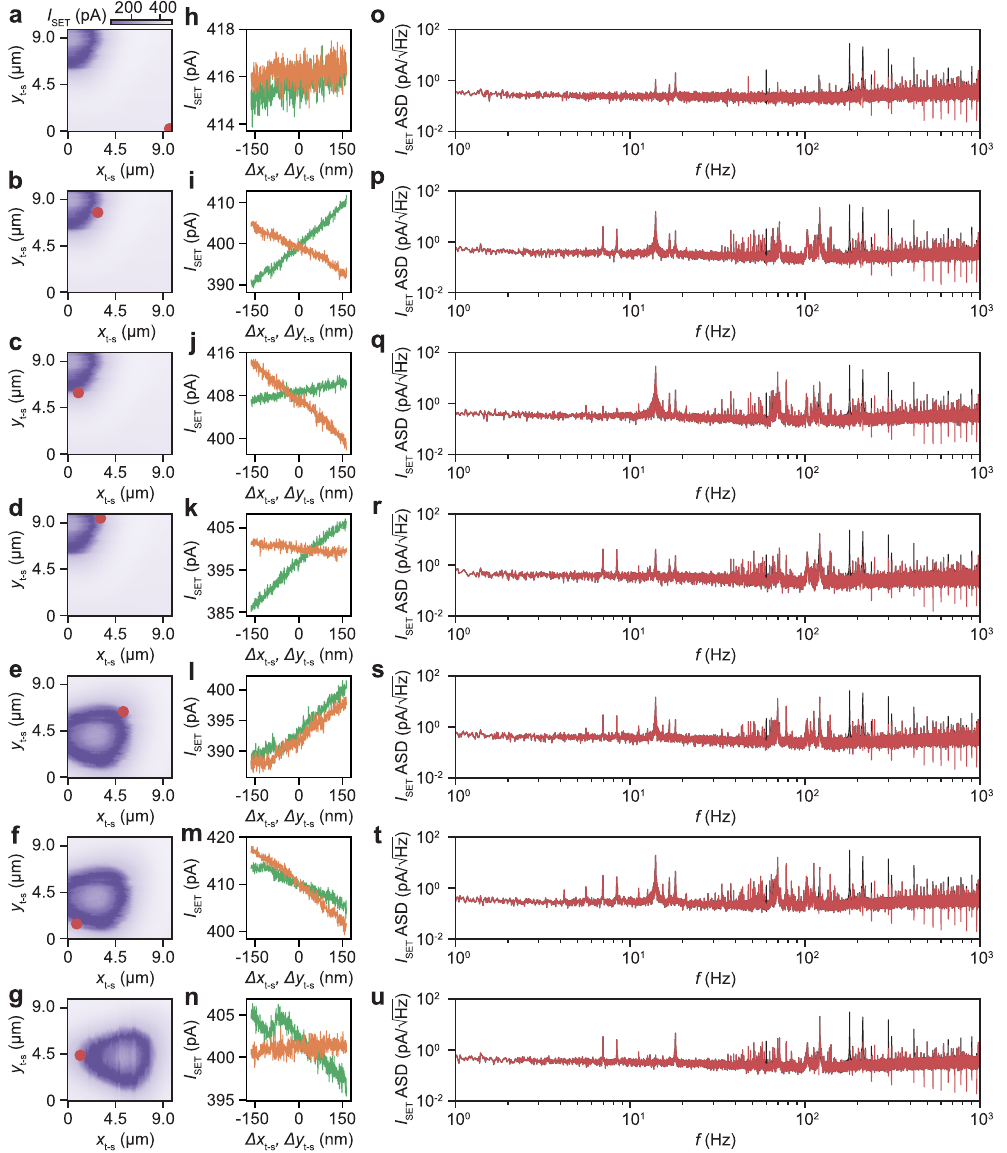}
\caption{Full dataset of $xy$ vibration measurements with the SET.
Each row corresponds to measurements from a given spatial location.
\textbf{a--g}~Maps of $I_{\text{SET}}(x,y)$ near a patterned feature on the test sample.
The red dot in each map marks the position where the vibration data are measured.
\textbf{h--n}~$I_{\text{SET}}$ at local flat band gate voltage measured at locations in \textbf{a--g} along the trajectories in the $x$- (green) and $y$- (orange) directions.
\textbf{o--u}~$I_{\text{SET}}$ noise spectra measured at the locations marked in \textbf{a--g}.
The raw spectra are shown in black, and incorporating digital notch filters applied to the raw data yields the red spectra.}\label{figS7}
\end{figure}

Seven spectra measured at different $xy$ locations are used to quantify the $x$ and $y$ tip-sample vibrations (Fig.~\ref{figS7}).
In general, at a given location, contributions from both $x$ and $y$ vibrations will mix into the raw data if the local electrostatic potential has a slope in both directions.
To account for this, we measure the total current noise spectra at each location, as well as the SET current gradients along each cardinal direction.
By measuring near the local flat band voltage, we eliminate the $z$ coupling in the spectra, which reduces Eq.~3 in the main text to
\begin{equation}
\delta I_{\text{SET}}^2 = \delta I_{\text{SET,0}}^2 + \left(\delta x_{\text{t-s}}\frac{\partial I_{\text{SET}}}{\partial x} + \delta y_{\text{t-s}}\frac{\partial I_{\text{SET}}}{\partial y}\right)^2.\label{eq:SET_coupled_xynoise}
\end{equation}
However, as discussed in Section~\ref{subsec_corr} above, the dominant modes in our $x$ and $y$ vibrations occur at distinct frequencies.
Therefore, we can further approximate the total current noise with $x$ and $y$ vibrations contributing independently:
\begin{equation}
\delta I_{\text{SET}}^2 = \delta I_{\text{SET,0}}^2 + \delta x_{\text{t-s}}^2\left(\frac{\partial I_{\text{SET}}}{\partial x}\right)^2 + \delta y_{\text{t-s}}^2\left(\frac{\partial I_{\text{SET}}}{\partial y}\right)^2.\label{eq:SET_xynoise}
\end{equation}
We then implement a non-negative least-square fit procedure using Eq.~\ref{eq:SET_xynoise} to extract the $x$ and $y$ vibrations based on data from all seven locations.
As the $xy$ spectra are also affected by fluctuations in the non-vibrational noise, we again filter them prior to the fitting process.
The fit takes the current noise power spectra density at a specific frequency $\delta I_{\text{SET}}^2$ as the dependent variable, squared current gradients $(\partial I_{\text{SET}}/\partial x)^2$ and $(\partial I_{\text{SET}}/\partial y)^2$ as two independent variables, and we implement non-negative least-squares fitting.
We perform this fitting at every frequency, and thus obtain the fitted coefficients $\delta x_{\text{t-s}}^2$ and $\delta y_{\text{t-s}}^2$ as power spectra of $x$ and $y$ vibrations.
Integrated $x$ and $y$ vibrations from 1 Hz to a given frequency $f$ are calculated from:
\begin{equation}
\sigma_{x_{\text{t-s}}} = \sqrt{\int_{1\text{~Hz}}^{f} \delta x_{\text{t-s}}^2 df}
\hspace{20pt}
\text{and}
\hspace{20pt}
\sigma_{y_{\text{t-s}}} = \sqrt{\int_{1\text{~Hz}}^{f} \delta y_{\text{t-s}}^2 df}.\label{eq:SET_xy_ts}
\end{equation}

\begin{figure}[t]
\centering
\includegraphics{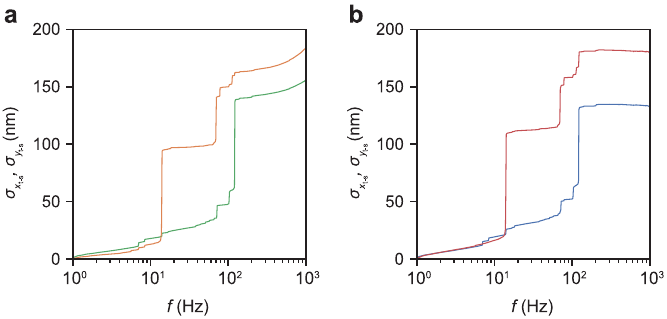}
\caption{\textbf{a}~Fitted $x$ (green) and $y$ (orange) vibrations based on fitting all spectra shown in Fig.~\ref{figS7}.
\textbf{b}~Integrated $x$ (blue) and $y$ (red) vibrations from the $x$- and $y$-dominant spectra corresponding to the blue and red curves in Fig.~6c.
We assume the vibrational contributions in the $x$- or $y$-dominant spectrum are from $x$ or $y$ vibrations only, and integrate the power spectra of $x$ or $y$ vibrations obtained from Eq.~\ref{eq:SET_xynoise}.}\label{figS8}
\end{figure}

We obtain RMS vibrations up to 300~Hz of 143~$\pm~13$~nm and 166~$\pm~15$~nm for $x$ and $y$, respectively.
These values are similar to estimates from directly integrating individual spectra domainated by $x$- or $y$-vibrations (see Fig.~\ref{figS8}).
Although we plot data up to 1~kHz, data at the highest frequencies (300~Hz to 1~kHz) exhibit unphysical artifacts in both panels of Fig.~\ref{figS8}, which we explain below.
Over this frequency range the spectra are dominated by a featureless non-vibrational background, which should be identical for all spectra.
One would therefore expect the fitted $\delta x_{\text{t-s}}^2$ and $\delta y_{\text{t-s}}^2$ to be zero in this frequency range, or, considering random fluctuations of power spectral density, to be fluctuating about zero.
However, our non-negative fitting algorithm prohibits the fitted coefficients to be less than zero, generating a positive white noise in the fitted $\delta x_{\text{t-s}}^2$ and $\delta y_{\text{t-s}}^2$ spectra.
After integration over a relatively wide bandwidth, an upturn, which is an artifact from this non-negative fitting, appears at higher frequencies in Fig.~\ref{figS8}a.
In contrast, for Fig.~\ref{figS8}b, a slight decrease of integrated vibrations can be seen in this frequency range, which we attribute to a slight drift of background noise among different spectra.

We also comment that while the maximum vibration amplitude for the corresponding highest RMS amplitude we have obtained is larger than the measured spatial range of $I_{\text{SET}}$ (around $\pm$150~nm) about each location, the notch-filtered current time series are symmetric about the current setpoint.
This indicates the current response is still linear over the full range of vibration amplitudes because beyond the linear regime, the shape of the SET Coulomb blockade oscillations is such that the nonlinearity has different magnitude on either side of the current set point.

\clearpage

\section{Additional SET Coulomb blockade thermometry measurements}

\begin{figure}[h]
\centering
\includegraphics{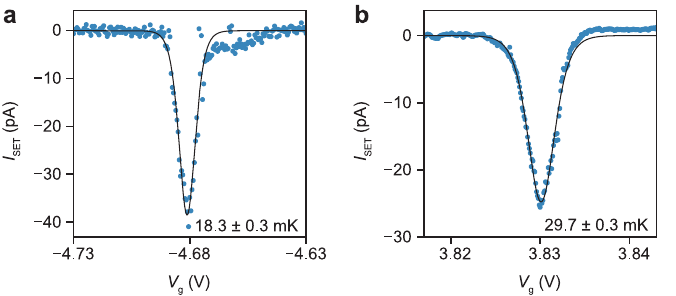}
\caption{\textbf{a}~Line cut of $I_{\text{SET}}$ as a function of $V_{\text{g}}$ near $V_{\text{g}} = -$4.68~V at SET source-drain bias $V_{\text{SET}}$ = $-$0.58~mV, slightly above the superconducting gap.
The electron temperature is fitted as 18.3~$\pm$~0.3~mK.
The SET tip is the same one described in main text.
\textbf{b}~Line cut of $I_{\text{SET}}$ as a function of $V_{\text{g}}$ near $V_{\text{g}} =$~3.83~V at $V_{\textrm{SET}}$ = $-$0.618~mV for another SET tip.
The electron temperature is fitted as 29.7~$\pm$~0.3~mK.}\label{figS9}
\end{figure}

Here we present additional SET Coulomb blockade thermometry measurements.
Figure~\ref{figS9}a shows a line cut of $I_{\text{SET}}$ at the apex of another Coulomb blockade diamond from the same tip at $V_{\text{g}} = -$4.68~V (see Fig.~7b in the main text for the Coulomb diamond).
The electron temperature fitted from this peak is 18.3~$\pm$~0.3~mK, similar to that obtained in the main text.
We also show in Fig.~\ref{figS9}b another example of Coulomb blockade thermometry from a different SET tip.
Here, $I_{\textrm{SET}}$ is measured near $V_{\text{g}} =$~3.83~V, at the apex of a Coulomb blockade diamond for this tip; fitting to the standard expression yields an electron temperature 29.7~$\pm$~0.3~mK.

The quoted $<$1~mK uncertainties for the above fits only reflect the uncertainties from the fitting procedure, and the range of fitted temperatures likely provides a more accurate estimate of the variation in electron temperature than the fitting uncertainties.
We note that including the shoulder in the fitting yields similar temperatures, and values within the above range are obtained at other (slightly) higher source-drain biases as well.
The range of different electron temperatures from the above measurements likely reflect slight differences in thermalization of different SET tips and random fluctuations of local charge traps for different Coulomb blockade peaks in the same tip.

\end{document}